\documentclass[12pt]{iopart}

\usepackage{graphicx}
\usepackage{verbatim}
\usepackage{amssymb}

\usepackage[english]{babel}
\usepackage{color}

\expandafter\let\csname equation*\endcsname\relax

\expandafter\let\csname endequation*\endcsname\relax

\usepackage{url}
\usepackage{graphicx}
\usepackage{graphics}
\usepackage{multicol}
\usepackage{footmisc}
\usepackage{amssymb}
\usepackage{amsmath}
\usepackage{algorithm,algpseudocode}

\usepackage{iopams}  

\begin{document}

\title[]{Bayesian Model Parameter Learning in Linear Inverse Problems: Application in EEG Focal Source Imaging}

\author{Alexandra Koulouri$^{1,2}$ and Ville Rimpil\"ainen$^{3,4}$}

\address{$^1$ Tampere University, Centre of Mathematics, Faculty of Information Technology and
Communication Sciences, Tampere, Finland}
\address{$^2$ University of Bath, Institute for Mathematical Innovation, Bath, United Kingdom}
\address{$^3$ University of Liverpool, Department of Physics, Liverpool, United Kingdom}
\address{$^4$ University of Bristol,
School of Physics, Bristol, United Kingdom}
\ead{a.koulouri84@gmail.com, vrimpila@gmail.com}

\begin{abstract}
Inverse problems are often described as limited-data problems in which
the signal of interest cannot be observed directly. Therefore, a physics-based forward model that relates the signal with the observations is typically needed. Unfortunately, unknown model parameters and imperfect forward models can still undermine the signal recovery. Even though supervised machine learning techniques offer promising avenues to improve the robustness of the solutions, we have to rely on model-based learning when there is no access to ground truth for the training. In this work, we studied a linear inverse problem that included an unknown non-linearly related model parameter and utilized a Bayesian model-based learning approach that allowed reliable signal recovery and subsequently estimation of the unknown model parameter. This approach, often referred to as Bayesian Approximation Error approach, employed a simplified model of the physics of the problem augmented with an approximation error term that compensated for the simplification. An error subspace was spanned with the help of the eigenvectors of the approximation error covariance matrix which allowed, alongside the primary signal, simultaneous
estimation of the induced error. The estimated error and signal were then used to determine the unknown model parameter. For the model parameter estimation, we tested several different approaches: a conditional Gaussian regression, an iterative (model-based) optimization, and a Gaussian process that was modeled with the help of
physics-informed learning. In addition, alternating optimization was used as a reference method. As an example application, we focused on the problem of reconstructing brain activity from EEG recordings (a.k.a. EEG source imaging) under the condition that the electrical conductivity of the patient’s skull was unknown in the model. Poorly selected conductivity values cause well-documented artifacts in the EEG source imaging results, and the determination of patient-specific head tissue conductivities is a significant technical problem. Our results demonstrated clear improvements in EEG source localization
accuracy and provided feasible estimates for the unknown model parameter, skull conductivity.

\end{abstract}

%
\vspace{2pc}

\section{Introduction}

\subsection{Model Uncertainties in Inverse Problems}

In many biomedical imaging, seismic imaging, remote sensing, and tomographic applications, the features of interest cannot be observed directly but must be inferred from other measurable quantities. This is known as an inverse problem \cite{Kirsch1996}. The limited, complicated, or distorted nature of observations, coupled with inherent limitations such as large null spaces in model operators, often make inverse problems ill-posed \cite{Tikhonov1943}. To address this, standard regularization techniques are often used \cite{Plato1990,Engl1996,Hansen1998}. Strategies to learn effective regularizers directly from the data are a promising field of research \cite{Benning2018,Afkham2021,Mattsson2023,Burger2023}. However, reliable and accurate computational models that relate the features of interest and the observations are also needed. 
Unfortunately, uncertainties are usually present in these models due to the modeling formulation itself (such as discretization that is required to achieve numerical solutions), imperfect sensing, or lack of knowledge of suitable model parameter values, which can significantly affect the final solution.

In applications like computerized tomography and magnetic resonance imaging \cite{Campisi2007,Ni2024}, well-established methods are employed to tackle unknown model parameters. In particular, for the estimation of signals of interest and model parameters, the corresponding bilinear optimization problems can be approached by convexifying the problem using methods such as linearization (lifting) \cite{Ahmed2014,Ling2015,Ni2024} or by alternating minimization algorithms \cite{Bolte2010,Li2019}. However, optimizing both the primary unknowns and the model parameters can be time consuming, convergence is not always guaranteed 
or several reinitializations may be required, and approximations for parameters that are {\it deep} in the model may not always work. 
Techniques that handle model uncertainties with the help of machine learning (ML) and data-driven approaches \cite{Adler2017,Adler2018,Jin2017} represent a new direction to accelerate computations and improve estimates in inverse problems \cite{Arridge2019,Lunz2021,Arridge2023}. However, unlike in many image and pattern recognition problems \cite{LeCun2015,Murphy2022}, in inverse problems there is often a limited or no access to ground truth data to train ML algorithms and deep neural networks \cite{Campisi2007,Murphy2022}. In this study, we revisit the application of machine learning to severely ill-posed inverse problems when no access to ground truth is possible. Specifically, we generate training data stemming from physics-based models and integrate this domain-specific knowledge into the inverse problem formulation and recover both the primary unknown and extract an estimate for the (unknown) model parameter. We approach the problem from a Bayesian perspective, which facilitates the
design of optimization algorithms under model uncertainty \cite{Kaipio,Kaipio2007}. 

As an application, we will focus on imaging brain sources from EEG recordings, due to its importance as a low-cost diagnostic tool and the inherent mathematical challenges and associated model uncertainties. In particular, we study imaging under the influence of erroneous skull conductivity (too high or too low), which has been shown to significantly affect the imaging solution \cite{Vanrumste2000,dan11,lew13,Ollikainen1999,mon14,Rimpilaeinen2019,Vorwerk2019}. { In numerical experiments, we used simulated EEG recordings generated by single dipole sources in the somatosensory cortex of the brain, as similar data sets (EEG and MEG) have previously been found to be suitable for skull conductivity calibration \cite{Lew09b,Papageorgakis2017,fuc98,hua07,ayd14,Antonakakis2020}. }Subsequently, the {inferred }conductivity value could be used in further EEG studies and in the optimization/planning of transcranial electrical brain stimulation treatment \cite{Schmidt2015,SATURNINO2019821,McCann_2021}. We remark that the developments of this paper are not limited to EEG source imaging but can also be used, for example, in other imaging and deconvolution problems.

\subsection{Bayesian Approximation Error Approach and Our Contributions}
{In this work, we build on the growing use of Bayesian statistics in machine learning by providing methodological approaches to efficiently solve linear inverse problems with model parameter uncertainties.} 
In particular, our developments are based on the well-known 
Bayesian approximation error (BAE) approach \cite{Kaipio,kaipio2013} in which a \emph{standard} (approximate) observation model is used in the imaging accompanied with an additive error term that takes into account the effects of the approximation. 
Similarly as in \cite{Nissinen2011a,kaipio2013}, we first derive a posterior distribution which simultaneously predicts the primary unknown signal and a low-order representation of the model-induced error. For the representation, we employ the top-eigenvectors of the modeling error covariance, obtained through Monte Carlo simulations. In addition to this, we derive an analytical expression for the modeling error covariance to investigate the connections between the error, model parameter, and primary signal. The provided insight 
allowed us to propose other Bayesian strategies to infer the model parameter. 

In the application domain, we evolve our previous work, in which we demonstrated that BAE can improve source localization when skull conductivity is unknown \cite{Rimpilaeinen2019}, and our preliminary study on simultaneous estimation of focal source activity and skull conductivity from EEG recordings \cite{Koulouri2020}. 
Here, with the help of simulated EEG data that corresponded to
focal source activity in the somatosensory brain area, we demonstrate the potential of our approach
to reconstruct the underlying focal sources and low-order estimates for the errors
induced by unknown skull conductivity. Subsequently, these estimates
are used to approximate the skull conductivity.

Our contributions are twofold. First, we propose a Bayesian model-based learning technique for linear inverse problems with unknown model parameters; second, we present different ways to estimate both the primary unknown and the (unknown) model parameter. We demonstrate the approach by imaging the source configuration (primary unknown) of the EEG brain imaging problem and estimating the unknown skull conductivity (model parameter) in the observation model. 

Although the BAE approach has been studied in a variety of settings \cite{Lipponen2011,Nissinen2009,Arridge2006,Kolehmainen2009,Tarvainen2010a,kou16,Mozumder2013,Alexanderian_2024,Candiani_2021,Nicholson_2023}, its use in the estimation of model parameters has been rather limited \cite{Nissinen2011a}. Therefore, in this paper, we concentrate on the model parameter learning aspect of BAE, offer insight on the conditions when it works, and propose two new ways to improve the model parameter estimates. Namely, we first propose to replace the previously used linear regression, that is based on a conditional Gaussian (CG) approximation \cite{Nissinen2011a}, with an iterative algorithm, and second, we propose to use a Gaussian process (GP) to infer the unknown model parameter. We show that these new developments are superior to the CG as the boundary data depend non-linearly on the model parameter.
Finally, we compare our methodology with other standard approaches dealing with blind inverse problems (in which model parameters are partly or fully unknown) and show that the proposed methodologies, which only rely on precomputed training data and off-line statistical learning, can outperform alternating optimization methods when unknown parameters are deep in the model. 

{The paper is organized as follows. We first provide mathematical preliminaries in Section 2. In Section 3, we then present a general overview of the Bayesian Approximation Error (BAE) approach and how it can be exploited for model parameter estimation. In Section 4, we describe how the BAE and the proposed model parameter estimation techniques can be applied to a model problem in which the source activity of the brain (primary unknown) and the electrical conductivity of the human skull (model parameter) are estimated from EEG recordings. Section 5 describes the implementation of the numerical simulations followed by the results and discussion. Section 6 consists of conclusions and future work. Finally, the appendices provide further details on the theory and analysis and a step-by-step pipeline on how to implement the BAE approach with model parameter estimation.}

\section{Background}
\label{sec:theory}

\subsection{Bayesian Inference in Linear Inverse Problems}

We consider the discrete observation model
\begin{equation}
\label{eq:obm2}
v = A(\sigma) x + e,
\end{equation}
where $v \in \mathbb{R}^m$ denotes the measurements, $m$ is the number
of measurements, $A(\sigma) \in \mathbb{R}^{m \times n}$ ({$m\ll n$}) is the
system matrix (discrete forward model) that depends on a model parameter 
$\sigma\in\mathbb{R}^{N_{\sigma}}$, $x\in \mathbb{R}^{n}$ is the distributed
signal (e.g. amplitudes in $n$ locations) that we aim to solve. Finally, $e$ denotes the measurement noise that is modeled as Gaussian $e\sim \mathcal{N}(e_*,\Gamma_e)$ with mean $e_*$ and covariance $\Gamma_e$. The model parameter $\sigma$ is usually (unrealistically) considered known. However, in this paper, we treat $\sigma$ as unknown, and subsequently estimate it alongside our \emph{primary} unknown $x$.

In the Bayesian framework, the solution to the inverse
problem is the posterior density
\begin{equation}\label{post}
\pi(x|v) \propto \pi(v|x)\pi(x),
\end{equation}
where $\pi(v|x)$ is the likelihood and $\pi(x)$ the prior.

In this case, the likelihood can be
formulated as
\begin{equation}\label{lik1}
\pi(v|x) \propto
\exp\Big(-\frac{1}{2}(v-A(\sigma)x-e_{*})^\mathrm{T}\Gamma_e^{-1}(v-A(\sigma)x-e_{*})\Big).
\end{equation}

The model parameter $\sigma$
is usually unknown and we want to obtain information on it, in addition to $x$. These problems are often called blind
inverse problems. Such problems are much harder to solve than standard inverse problems. One approach is to
estimate the parameters of the model and the signal by alternating optimizations, but this has certain limitations; namely, convergence is not always reached, and the computations can be time-consuming. Ideally, we would like to marginalize out the unknown 
parameter $\sigma$; however this is not usually tractable. Hence, instead of the accurate matrix
$A(\sigma)$, an approximate (standard) matrix $A_{0} = A(\sigma_0)$ with fixed parameters $\sigma_0$ is often used in the inversion.

\subsection{Bayesian Machine Learning}

Machine learning in inverse problems relies on training a model with known data, $D=\{ (x_s,v_s)\}_{s=1:S}$, in order to predict underlying signals $x$ given a new set of observations, $v$. In Bayesian ML, model parameter $\sigma$ is a random variable with distribution $\pi(\sigma|D)$, {$x$ and $\sigma$} are statistically independent and the likelihood is expressed as 
 \begin{equation}\label{eq:likelhood}
     \pi(v|x) =\int \pi(v|x,e, \sigma)\pi(e)\pi(\sigma|D)d\sigma de.
 \end{equation}
To estimate the primary unknown $x$, we have to estimate the previous likelihood; however, marginalizing out directly the unknown $\sigma$ parameter is not always tractable (especially for non-linear parameters). Even if we approximate $\pi(\sigma|D)=\delta(\sigma-\hat{\sigma})$, the question on how we estimate $\hat{\sigma}$ still remains open. Furthermore, in such inverse problems where the access to ground truth distribution for $x$ is (most of the time) impossible, we have to rely on training techniques that employ model-based learning and sampling distributions. However, to guarantee that feasible model parameters have been selected, the training data $D$ (obtained from model distributions) has to be produced under carefully designed simulations.

\section{Bayesian Approximation Error Approach}

\subsection{Overview} \label{sec:overv}

To ease the model parameter estimation, we rewrite the observation model
(\ref{eq:obm2}) with the help of an approximate model $A_0 = A(\sigma_0)\in \mathbb{R}^{m\times n}$, where $\sigma_0$ has fixed values, and remove the non-linearity with respect to the { unknown model parameter $\sigma$} as 
\begin{equation}\label{eq:AMEA}
v= A_{0}x+\varepsilon+e,
\end{equation}
where the additive error term $\varepsilon \in \mathbb{R}^m$, referred to as {\em approximation error} \cite{Kaipio,kaipio2013}, accounts for the simplification. In particular, the approximation error term is related to the unknown parameter as
\begin{equation}\label{eq:approximationerror}
\varepsilon(x,\sigma) = A(\sigma) x-A_0x. 
\end{equation} 
Furthermore, when the variability of the error term $\varepsilon$ is relatively low, it can be expressed as a linear combination of few basis functions, for example, as in \cite{Nissinen2011a,kaipio2013} by using the eigenvalue decomposition of the error covariance, $\Gamma_{\varepsilon} = \Sigma_{k=1}^{m} \lambda_k w_k w_k^{\mathrm{T}}$, where $\lambda_k>0$ (in descending order) are the eigenvalues and $w_k \in \mathbb{R}^m$ the eigenvectors. Now, we can write the approximation error as 
\begin{equation}\label{eq:ae}
\varepsilon(x,\sigma)=\varepsilon_*+\varepsilon'+\varepsilon''=W^p\alpha+\varepsilon_*+\varepsilon'',
\end{equation}
{
where $\varepsilon_*=\mathbb{E}[\varepsilon]$ is the statistical mean of $\varepsilon$ and $W^p=[w_1,\ldots,w_p]\in\mathbb{R}^{m\times p}$ ($m>p$) are the $p$ top eigenvectors.  
The term $\varepsilon'=W^p\alpha$ represents the dominant part of the variability in the error,
while $\varepsilon''$ is a less informative term described by the remaining $m-p$ eigenvectors with mean $\mathbb{E}[\varepsilon'']=0$ and covariance $
\Gamma_{\varepsilon''} = \Sigma_{j=p+1}^{m}\lambda_j w_j
w_j^{\mathrm{T}}$.  The coefficient vector $\alpha\in\mathbb{R}^p$ represents the low-dimensional error components with mean $\mathbb{E}[\alpha]=0$ and diagonal covariance $\mathrm{cov}[\alpha]=\mathrm{diag}\{\lambda_1,\ldots,\lambda_p\}$.  More details on this error decomposition are given in Appendix A1. 
Now, by substituting the decomposed error into the observation model (\ref{eq:AMEA}) we get
\begin{equation}
v = A_0 x + \varepsilon_* + W^p \alpha + \varepsilon'' + e.
\label{eq:observationmodel}
\end{equation}
In practical applications, the basis $W^p$ and corresponding statistics can be learned from training data sampled from the joint model distributions $\pi(x)$ and $\pi(\sigma)$, or computed semi-analytically \cite{Nissinen2011a,kaipio2013}. Then, we can
infer simultaneously $x$ and $\alpha$ based on Bayes' rule, $\pi(x,\alpha|v) \propto \pi(v|x,\alpha) \pi(x,\alpha)$.
 } 

Consequently, one way to obtain parameter estimates is by maximizing the conditional probability distribution $\hat{\sigma}:=\max_{\sigma}\pi(\sigma|\alpha)$,  considering a joint Gaussian for $(\sigma,\alpha)$ as has been done previously \cite{Nissinen2011a}. 
However, this approach has limitations, since it considers that $\sigma$ and $\alpha$ are sensibly related through a linear relationship (or at least monotonic relationship), which may not be valid, and the approach also ignores the fact that $\alpha$ depends on the signal $x$. This approximation can work well when the difference $\sigma-\sigma_0$ is small. 

In a more general case, 
{
we can consider a mapping from  $(\alpha,x)\rightarrow\sigma$. Since the analytical expression of this mapping is often unknown (or too complicated to be derived), we can use non-parametric methods, such as a Gaussian process (GP), for the inference of the model parameter $\sigma$. Since $x$ and $\sigma$ are usually independent (this depends on the underlying physics of the problem), the only way $x$ can affect $\sigma$ in this model is through its interaction with the variable $\alpha$; in our case, this kind of interaction occurs because $\alpha$ depends on $x$ through equations (\ref{eq:approximationerror}) and (\ref{eq:ae}). Hence, GP learns a statistical relationship, not a causal one. GP is effectively learning the inverse function of $\sigma$ from the structure imposed by $A(\sigma)$. Overall, as we shall see later by approximating (\ref{eq:approximationerror}) and obtaining statistical knowledge through training data, we can recover meaningful estimates $\hat{\sigma}$ under certain conditions.}

\subsection{Bayesian Inference of Primary Signal and Approximation Error}
To jointly estimate both $x$ and $\alpha$, we need the posterior distribution $\pi(x, \alpha \mid v)$. To reduce computational complexity, we adopt an approximation in which the variables $(x, \alpha, e, \varepsilon'')$ are treated as mutually Gaussian and uncorrelated\footnote{We note that even though in reality approximation errors usually depend on $x$, this \emph{technical} approximation often results in very similar inverse solutions \cite{Kaipio,Kaipio2007,kaipio2013}.} \cite{Nissinen2011a,kaipio2013}. The resulting approximate likelihood is
\begin{eqnarray}\label{lik2}
\tilde{\pi}(v | x, \alpha) \propto \exp
\Big( && -\frac{1}{2}(v-A_{0}x-\varepsilon_* -W^p\alpha-e_{*})^\mathrm{T} \\\nonumber 
&& (\Gamma_{\varepsilon''}+\Gamma_e)^{-1}(v-A_{0}x-\varepsilon_{*}-W^p\alpha -e_*)\Big).
\end{eqnarray}
The corresponding posterior density is
\begin{equation}\label{eq:posteriorBAE}
\tilde{\pi}(x,\alpha|v) \propto \mathrm{exp}
\left(-\frac{1}2\|L_{\varepsilon''+e} \left( v-A_0x-W^p\alpha-\varepsilon_*-e_*\right) \|_2^2\right)
\;\pi(x,\alpha),
\end{equation}
where $L_{\varepsilon''+e}$ comes from the Cholesky factorization of
$(\Gamma_\varepsilon+\Gamma_e)^{-1}$.
Finally, the maximum a posteriori (MAP) estimate is obtained by solving:
\begin{equation}\label{joint_inference}
(\hat{x}, \hat{\alpha}) \leftarrow 
\arg\min_{x, \alpha} \left\{
\frac{1}{2} \left\| L_{\varepsilon''+e} 
\left(A_0 x + W^p \alpha + \varepsilon_* + e_* - v \right) 
\right\|_2^2 
- \log \pi(x, \alpha)
\right\}.
\end{equation}
{Next, we can use the point estimates $(\hat{x},\hat{\alpha})$ to infer $\sigma$ with the help of learning techniques}.

\subsection{Inference of the Model Parameter}

\subsubsection{Variable Interactions} \label{sec:interact}
{
The inference of the model parameter \( \sigma \) from the estimated error  (as in \cite{Nissinen2011a}) 
\begin{equation}\hat{\varepsilon}\approx W^p\hat{\alpha} + \varepsilon_* \end{equation} is feasible if the basis functions \( W^p \) primarily capture the error induced by the unknown model parameter \( \sigma \). However, this is not always the case, and in many situations, the estimated signal \( \hat{x} \) must also be considered when estimating \( \sigma \).  Without loss of generality in the following analysis we consider that $\sigma>0$ is a scalar. 

In particular, even though \( x \) and \( \sigma \) are independent variables, \( x \) can still influence \( \sigma \) through its interaction with \( \alpha \). This implies that we are dealing with an \textbf{interaction effect} rather than a direct causal relationship.  To better understand how the linear and nonlinear effects of \( \sigma \) and \( x \) are incorporated into the basis functions \( W^p \) and, consequently, affect the estimation of \( \alpha \) (which we use to infer \( \sigma \)), we express the error covariance semi-analytically using a Taylor expansion of the error (\ref{eq:approximationerror}).  

Expanding around the mean \( \sigma_* \), we obtain:  
\begin{equation}\label{eq:TaylorExpansion}
\varepsilon(x,\sigma) = A(\sigma_*)x - A_0x + \sum_{k} J_k(\sigma_*) z_k(\sigma) x,
\end{equation}  
where  
$
J_k = \frac{1}{k!} \frac{\partial^k A(\sigma)}{\partial \sigma^k} \Big|_{\sigma = \sigma_*} \in \mathbb{R}^{m \times n}$ and $ z_k(\sigma) = (\sigma - \sigma_*)^k$.

For simplicity, we assume \( \sigma_* = \sigma_0 \), and the error covariance can be expressed as:  
\begin{equation}
\Gamma_\varepsilon = \sum_k J_k \, \operatorname{cov}[x z_k] \, J_k^\mathrm{T} + \sum_{k \ne j} J_k \, \operatorname{cov}[x {z_k}, x z_j^\mathrm{T}] \, J_j^\mathrm{T}.
\end{equation}
Since \( z_k \) and \( x \) are mutually independent, it follows from \cite{Bohrnstedt1969} that the covariance matrix of their product is given by:  
\begin{equation}
   \operatorname{cov}[x z_k] = \operatorname{cov}[x] \operatorname{var}[z_k] + \operatorname{cov}[x] \mathbb{E}[z_k]^2 + \operatorname{var}[z_k] \mathbb{E}[x] \mathbb{E}[x]^\mathrm{T}.
\end{equation}  

Furthermore, when the mean $\mathbb{E}[x]=0$ and $x$ has isotropic covariance, i.e. $\mathrm{cov}[x]=\gamma I_n$, we get
\begin{equation}
\Gamma_\varepsilon = \gamma \sum_k J_k \left( \operatorname{var}[z_k] + \mathbb{E}[z_k]^2 \right) J_k^\mathrm{T}
+ \gamma \sum_{k \ne j} J_k  \mathbb{E}[z_k z_j^\mathrm{T}] J_j^\mathrm{T}.
\end{equation}
This expression for the error covariance suggests 
that \( x \) has primarily a \textbf{scaling effect} on the error covariance. 
Furthermore, we need to determine whether matrix \( J_k \) implicitly encodes information about \( x \), potentially arising from the modeling process (e.g., linearization of a nonlinear problem).
Overall, we expect the basis functions \( W^p \) (obtained via the eigenvalue decomposition of \( \Gamma_\varepsilon \)) to capture error contributions depending both from the unknown model parameter \( \sigma \) and the primary signal \( x \)\footnote{If $\sigma \,\in\mathbb{R}^{N_{\sigma}}$ where $N_\sigma>1$, we can obtain a similar expression, i.e. $
\Gamma_\varepsilon = \gamma\sum_{k,j} J_k \left( \operatorname{cov}[z_k, z_j] \otimes I_n \right) J_j^\top.
$, where $J_k\,\in\,\mathbb{R}^{m\times (n \cdot N_\sigma)}$ and $\otimes$ is the Kronecker product.}. 
 }

\subsubsection{Conditional Gaussian {(CG)} Approximation}

In the Bayesian framework, we expect to obtain model parameter estimates  by maximizing the conditional probability distribution
$\hat{\sigma}:=\max_{\sigma}\pi(\sigma|\alpha)$, 
{however this is not always tractable.}
Instead, a simple approach is to approximate $ \pi(\sigma|\alpha) \propto \pi(\sigma, \alpha)$ and rely on 
Monte Carlo simulations \cite{Nissinen2011a}. In particular, a joint Gaussian distribution for the pair ($\alpha,\sigma$) can be a good approximation for small differences $\sigma-\sigma_*$. { In other words,  first order Taylor approximation with only the linear term in (\ref{eq:TaylorExpansion}) is sufficient to describe the modeling error}. Therefore, given an estimate for the error coefficients $\hat{\alpha}$, the mean of $\pi(\sigma|\hat{\alpha})$ is given by 
\begin{equation}\label{conditional_conductivity_general} \hat{\sigma}_{*|\alpha} = \sigma_* + \Gamma_{\sigma \alpha}\Gamma_{\alpha}^{-1} \hat{\alpha},\end{equation}
where  $\sigma_*$ denotes the mean value of the postulated model parameter distribution, the cross-covariance $\Gamma_{\sigma \alpha}$ is estimated using samples of $\alpha^{(s)}$ and $\sigma^{(s)}$, and $\Gamma_\alpha=\mathrm{diag}\{\lambda_1,\ldots,\lambda_p\}$. {
As we shall see later (in Section \ref{sec:iterativealg}), in some cases the CG-based parameter estimation can be used to initialize a nonlinear iterative algorithm to improve the model parameter estimates. 
}
\subsubsection{Gaussian Process {(GP)} }
{
In more general cases, where $\alpha=g(x,\sigma)$, we can employ statistical techniques to learn the inverse function, $\sigma=f(\alpha,x)$, in order to predict $\sigma$.
As explained earlier (in Sections \ref{sec:overv} and \ref{sec:interact}), this expression does not mean that $x$ and $\sigma$ are dependent; it simply means that given $\alpha$, we may be able to refine our estimate of $\sigma$ by knowing $x$, as the value of $\alpha$ depends on both $\sigma$ and $x$.}
To infer $\sigma$, we can model $f$ to be distributed as a Gaussian process, $f\sim \mathcal{GP}(m,K)$, with
mean function 
$m(\alpha,x)=\mathbb{E}[f(\alpha,x)] $ and covariance (or kernel) function $K_{(\alpha,x)}=\mathbb{E}[(f(\alpha,x))-m(\alpha,x))((f(\alpha',x'))-m(\alpha',x'))^\mathrm{T}]$ \cite{Murphy2022}.
In practice, based on training input-output data, $\mathcal{D}=\{x^{(s)},\alpha^{(s)}\}_{s=1}^{N_s}$ and $f_\mathcal{D}=\{\sigma^{(s)}\}_{s=1}^{N_S}$ respectively, we can infer a new model parameter value $\hat{\sigma}$ given a new input estimated from (\ref{joint_inference}). Thus, we can predict $\hat{\sigma}$ given $\mathcal{\hat{D}}=(\hat{x},\hat{\alpha})$ and the training data $\mathcal{D}$.

The predictive distribution is $\pi(f_{\mathcal{\hat{D}}}|\mathcal{\hat{D}},\mathcal{D},f_\mathcal{D})\sim\mathcal{N}(\hat{f}_\mathcal{\hat{D}|\mathcal{D}},K_{\mathcal{\hat{D}}|\mathcal{D}})$. The conditional mean based on the training set is
\begin{equation}\label{eq:GP_inference1}
\hat{f}_\mathcal{\hat{D}|\mathcal{D}} =m(\hat{D})+K_{\mathcal{\hat{D}},\mathcal{{D}}}K_\mathcal{D}^{-1}(f_\mathcal{D}-m(\mathcal{D})),
\end{equation} 
where  
$K_{\mathcal{\hat{D}}|\mathcal{D}}= K_\mathcal{\hat{D}}-K_{\mathcal{\hat{D}},\mathcal{{D}}}K_\mathcal{D}^{-1}K_{\mathcal{\hat{D}},\mathcal{{D}}}^{\mathrm{T}}$ is the conditional variance, 
$K_{\mathcal{\hat{D}},\mathcal{{D}}}$ is a vector of covariances between every training case and $(\hat{\alpha},\hat{x})$, $K_\mathcal{D}$ is a matrix with the training data set covariance, and $K_\mathcal{\hat{D}}$ is the variance of $(\hat{\alpha},\hat{x})$. 
The choice of the mean function and kernel can be determined based on available (sample) data and properties (e.g. physics) of the inverse problem in question.

\section{{ Dipole Source Imaging and Skull Conductivity Estimation using Somatosensory EEG Data }}

{
In this section, we present a framework for joint estimation of brain source activity and skull conductivity using EEG responses. Building on the previously presented Bayesian approach, we present a modified single-dipole scanning technique to estimate single sources and approximate modeling errors. Then, we describe how the modeling error, which is dependent on the skull conductivity, can be exploited to calibrate the skull conductivity value. To achieve this, we use three strategies, a conditional Gaussian (CG) approximation, an iterative algorithm that can refine the CG-result (CG+Iter.), and finally a Gaussian process (GP) -based regression. In Section 5, we demonstrate the proposed approaches with simulated EEG recordings generated by single-dipole sources in the somatosensory cortex of the brain, as this kind of EEG (and MEG) datasets have been shown to be suitable for skull conductivity calibration  \cite{Lew09b,Papageorgakis2017,fuc98,ayd14,Antonakakis2020}.   
} 

\subsection{Simultaneous Approximation Error and Source Estimation using Model-based Learning and Single Dipole Scanning}

In this work, we apply the described Bayesian framework for the reconstruction of the unknown skull conductivity (model parameter) in the EEG source imaging problem alongside the source activity. Here, we consider the distributed source modeling \cite{Grech2008,Michel2019} where the primary unknown $x\in \mathbb{R}^{3n}$ is the distributed (electrical) current
dipole source configuration (or field) in a source space that consists of $n$ discrete locations, $\left| x_i \right| = \sqrt{ \mathrm{x}_{i1}^2 + \mathrm{x}_{i2}^2 + \mathrm{x}_{i3}^2 } $ is the amplitude of the source at location $i$, and $(\mathrm{x}_{i1},\mathrm{x}_{i2},\mathrm{x}_{i3})$ are the components of $x_i$ along the coordinate axes. The observation model is linear, as Equation (\ref{eq:obm2}), where
$v \in \mathbb{R}^m$ are the EEG recordings, and the system matrix $A(\sigma) \in \mathbb{R}^{m \times 3n}$ is called the \emph{leadfield} matrix, which depends on the model parameter, electrical conductivity of the skull
$\sigma\in\mathbb{R}^+$.

To obtain estimates for the source and the approximation error, we first compute the maximum a posteriori (MAP) estimates of the posterior (\ref{joint_inference}).
{ Since, in the numerical experiments (Section 5), we use simulated EEG recordings generated by single dipole sources within a restricted region (namely, the somatosensory cortex of the brain), we employ the dipole scanning solver that is widely used in EEG studies \cite{Fuchs1998,Rodriguez-Rivera2003,Hoeltershinken2024,knosche1997solutions}.} In the dipole scanning algorithm,
the main assumption is that only a single dipole source $x_i$ is active at a time. 

{
In Bayesian perspective, the standard dipole scanning algorithm solves $\hat{x}_i = \arg \max_{x_i}  \pi(v | x_i)$, where $\pi(v | x_i)$ is the likelihood considering a single source at location $i$, and the solution is the estimate $\hat{x}_i$  for which the likelihood $\pi(v | \hat{x}_i)$ is maximized. 

For the proposed framework, the approximate likelihood of the dipole scanning algorithm is 
\begin{equation}
\label{eq:ApproxLikelihoodDipoleScan}\tilde{\pi}(v|x_i,\alpha_i)\propto \exp{\frac{1}{2}\left( (v-A^i_0x_i+W^p_{i}\alpha_i+\varepsilon_{i*})^\mathrm{T}\Gamma_{e+\varepsilon_i''}^{-1} (v-A^i_0x_i+W^p_{i}\alpha_i+\varepsilon_{i*})\right)}.\end{equation}
Here the subscript $i$
denotes a specific dipole location, the superscript in $A^i(\sigma)$ denotes the
$i1$, $i2$, and $i3$ columns of the leadfield matrix and $\alpha_i\,\in\mathbb{R}^p$ are the error coefficients encoding the modelling error related to location $i$. 
In particular, the error related to location $i$ is $\varepsilon_i(\sigma)=A^i(\sigma)x_i-A_0^ix_i\,\in\mathbb{R}^m$ and its representation using basis functions is $\varepsilon_i(\sigma)=W^p_{i}\alpha_i+\varepsilon_{i*}+\varepsilon_i''$. 
The basis functions $W^p_i\,\in\mathbb{R}^{m\times p}$ are the $p$ top eigenvectors from the eigenvalue decomposition of the approximation error covariance matrix related to a source at location $i$ which is denoted by
$\Gamma_{\varepsilon_i} =\mathbb{E}
[(\varepsilon_i-\varepsilon_{i*})(\varepsilon_i-\varepsilon_{i*})^{\mathrm{T}}]\,\in\mathbb{R}^{m\times m}$
where $\varepsilon_{i*}\,\in\mathbb{R}^m$ is the mean approximation error at location
$i$.}

{
Therefore, 
 we can compute a MAP estimate of the pair $(\hat{x}_i,\hat{\alpha}_i)$ at each
location $i$, and then select as a solution the pair that results in the maximum posterior
 $\pi({x}_i,{\alpha}_i|v)\propto \pi(v|{x}_i,{\alpha}_i)\pi(x_i,{\alpha}_i)$. In particular, considering uniform prior for $\pi(x_i)$ and $\pi(\alpha_i)\sim \mathcal{N}(0,\Gamma_{\alpha_i})$ where $\Gamma_{\alpha_i}=\mathrm{diag}\{\lambda_{i1},\ldots,\lambda_{ip}\}$, we have the prior $\pi(x_i,\alpha_i)\propto\pi(\alpha_i)$. Then, the dipole scanning algorithm solves for $i=1,\ldots,n$
\begin{equation}\label{Map_per_location}
    (\hat{x}_i,\hat{\alpha}_i)=\arg \max_{x_i,\alpha_i}  \pi(v | x_i,\alpha_i)\,\pi(\alpha_i).
\end{equation}

\noindent
The final solution is
\begin{equation}
   (\hat{x}_l,\hat{\alpha}_l) \leftarrow\max_{i=1,\dots,n} \pi(v | \hat{x}_i,\hat{\alpha}_i)\,\pi(\hat{\alpha}_i), 
\end{equation}
where  index $l$ refers to the location that gives the maximum posterior. From (\ref{Map_per_location}), (\ref{eq:ApproxLikelihoodDipoleScan}) and for the given prior, we have that 
\begin{equation}\label{Map_per_location1}
    (\hat{x}_i,\hat{\alpha}_i)=\arg \max_{x_i,\alpha_i}  \exp{\left(-\frac{1}2 \left\|\begin{bmatrix}L_{\varepsilon^{''}_i+e} A_0^i & L_{\varepsilon^{''}_i+e}
W^p_{i}\\ \mathbf{0} & L_{\alpha_i}\end{bmatrix}\begin{bmatrix}
    x_i\\ \alpha_i
\end{bmatrix}-\begin{bmatrix}L_{\varepsilon^{''}_i+e} (v -\varepsilon_{i*}-e_*)\\
0\end{bmatrix}\right \|_2^2\right)},
\end{equation}
where $\|\cdot\|_2^2$ is the $\ell_2$-norm, $L_{\varepsilon^{''}_i+e}$ comes from the Cholesky factorization of $\Gamma_{\varepsilon_i''+e}^{-1}=(\Gamma_{\varepsilon_i^{''}}+\Gamma_e)^{-1}$ and $L_{\alpha_i}=\frac{1}2 \mathrm{diag}\{\lambda_{i1}^{-1/2},...,\lambda_{ip}^{-1/2}\}$ where $\Gamma_{\alpha_i}^{-1}=L_{\alpha_i}L_{\alpha_i}^\mathrm{T}$.
 }
Based on the previous, the MAP estimate for location $i$ is
\begin{equation}
\begin{bmatrix}
\hat{x}_i\\ \hat{\alpha}_i
\end{bmatrix}:=\begin{bmatrix}L_{\varepsilon^{''}_i+e} A_0^i & L_{\varepsilon^{''}_i+e}
W^p_{i}\\ \mathbf{0} & L_{\alpha_i}\end{bmatrix}^{-1} \begin{bmatrix}L_{\varepsilon^{''}_i+e} (v -\varepsilon_{i*}-e_*)\\
0\end{bmatrix}.
\end{equation}
Finally, the solution is the pair denoted with subscript $l$ that minimizes the functional
\begin{equation}\label{eq:Functional3}
(\hat{x}_l,\hat{\alpha}_l)\leftarrow\min_{i=1:n} \{\|L_{\varepsilon^{''}_i+e}(v-A_0^i
\hat{x}_i-\varepsilon_{i*}-W^p_{i}\hat{\alpha}_i-e_{*})\|_2^2+\|L_{\alpha_{i}}\hat{\alpha}_i\|_2^2
\}.
\end{equation}
Here, the corresponding primary error is given by $\hat{\varepsilon}'_l=W^p_{l}\hat{\alpha}_l$.

In the following subsections, we analyze the strategies for estimating $\sigma$ given the pair $(\hat{x}_l,\hat{\alpha}_l)$. 

\subsection{Skull Conductivity Estimation}
\label{sec:sce}

To obtain sensible estimates for the skull conductivity (model parameter), the estimated $\hat{\alpha}$ (and the corresponding modeling error) has to be clearly correlated with $\sigma$. To analyze this, we inspect first the covariance matrix of the {approximation error at location $l$ that is analytically given by
\begin{equation}\label{eq:covariance1Loc}
\Gamma_{\varepsilon_l}= 
\gamma \left(\mathrm{var}(\sigma)J_1^{l} {J_1^l}^\mathrm{T}+
\sum_{k,j\geq 2} \mathrm{cov}[z_k,z_j] J^l_k{J^l_j}^\mathrm{T}\right),
\end{equation}}where $z_k=(\sigma-\sigma_*)^k$,  $J_k^{l}=\frac{1}{k!}\frac{\partial^k A^l(\sigma_*)}{\partial
\sigma^k}\in\mathbb{R}^m$ is the $k$th derivative of the leadfield  $A^{l}(\sigma)$\footnote{In practice, we noticed that the derivatives of order $k,j \geq 2$ were negligible.} and $\gamma$ is the dipole variance. 
Here, without a loss of generality, we have considered that the standard conductivity is equal to the mean conductivity of the training data $\pi(\sigma)$, i.e. $\sigma_0\approx \sigma_*$, to ease our analysis.

From the previous expression for the error covariance, we observe that the source $x_l$ introduces a scaling effect on $\Gamma_{\varepsilon_l}$.
{ Because of that and since the error covariance is
$\Gamma_{\varepsilon_l}  \approx W^p_{l} \Gamma_{\alpha_l} (W^p_{l})^\mathrm{T}$,  
the bases $W_l^p$ 
are shaped primarily by the variability in the model parameter $\sigma$. However, due to the scaling effect of the source $x_l$ in $\Gamma_{\varepsilon_l}$, one can expect that the error coefficients $\alpha_l$ will be proportional to the amplitude of $x_l$.} Therefore, we can suspect that the methods that employ only the estimated error $\hat{\alpha}_l$ to infer $\sigma$, as the conditional Gaussian approximation described below, will not perform as well as the methods that include both, $\hat{\alpha}_l$ and the estimated source $\hat{x}_l$ as inputs. 

\subsubsection{Conditional Gaussian (CG) Approximation}

Given $\hat{\alpha}_l$, the conditional mean of the skull conductivity distribution $\pi(\sigma|\hat{\alpha}_l)$ is
\begin{equation}\label{conditional_conductivity} \hat{\sigma}_{\mathrm{CG}} = \sigma_* + \Gamma_{\sigma \alpha_l}\Gamma_{\alpha_l}^{-1} \hat{\alpha}_l,\end{equation}
where  $\sigma_*$ is the mean value of the postulated skull conductivity distribution,  $\Gamma_{\sigma \alpha_l}$ is the cross-covariance between $\sigma$ and $\alpha_l$  (estimated from the samples of $\alpha_l^{(s)}$ and $\sigma^{(s)}$, as will be described in Section~\ref{sec:MCMC}), and $\Gamma_{\alpha_l}=\mathrm{diag}\{\lambda_{1l},\ldots,\lambda_{pl}\}$ where $\lambda_{1p}$ are the first $p$ eigenvalues of covariance $\Gamma_{\varepsilon_l}$.

This Gaussian approximation approach can provide fast results, since it relies on off-line precomputed statistics based on training data. 
However, in our EEG source imaging case, the approximation error and the model parameter are non-linearly related, and the approximation error depends also on the dipole values.
Hence, the CG approximation may not give sufficiently accurate results further away from the linearization point $\sigma_*$. 

\subsubsection{Conditional Gaussian with Iterations (CG+Iter.)}
\label{sec:iterativealg}

In this case, it is possible to improve the CG-based model parameter estimates by iteratively updating the linearization point. Algorithm \ref{alg:LinearSkull} presents an iterative approach that uses the CG result as an initialization and also considers the estimated value of the source. 
\begin{algorithm*}[h!]\label{Algo:iter}
\caption{Iterative Linear Skull Conductivity Estimation} \label{alg:LinearSkull}
    \begin{algorithmic}
     \State 
     \textbf{Initialization}: initialize ${\sigma}_1=\sigma_{\mathrm{CG}}$ using the CG solution (\ref{conditional_conductivity}).\\
    
  \textbf{DO}
         Update leadfield model, $A^l(\sigma_{t})$\\
         \qquad Compute Jacobian matrix, $J_{\sigma_t}=\frac{\partial A^l(\sigma_{t})}{\partial \sigma_t}$ as in \cite{Vauhkonen1997} \\
         \qquad Compute, $b=w_{l1}\hat{\alpha}_l+\varepsilon_{l*}-(A^l(\sigma_t)-A_0^l)\hat{x}_l$\\
         \qquad Solve: $\sigma_{t+1}= \min_{\sigma} \vert\vert b - J_{\sigma_t} \hat{x}_l(\sigma-\sigma_t)\vert\vert_2^2$
              \\
 \textbf{WHILE $\vert\vert \sigma_{t+1}-\sigma_t \vert\vert_2 < \epsilon_{\mathrm{tolerance}}$}
    \end{algorithmic}
\end{algorithm*}

\subsubsection{Gaussian Process (GP)}

Another way to estimate model parameters is to employ a Gaussian process. In GP, we can  predict $\hat{\sigma}$ given  $\mathcal{\hat{D}}=(\hat{\alpha}_l,\hat{x}_l)$ as estimated from  (\ref{eq:Functional3}) and the training data $\mathcal{D}$. Hence, we have 
\begin{equation}\label{eq:GP_inference}
\hat{\sigma}_\mathrm{GP}=m(\hat{D})+K_{\mathcal{\hat{D}},\mathcal{{D}}}K_\mathcal{D}^{-1}(f_\mathcal{D}-m(\mathcal{D})).
\end{equation} 
Different kernel $K$ and mean functions $m(.)$ can model complicated relationships between variables, and the way they were chosen in this case is described in Section \ref{GP_functions}.

\section{{Implementation, Results and Discussion}}
{ In this Section, we use simulated EEG recordings generated by single-dipole sources within the somatosensory cortex of the brain to explore the capability of the proposed approaches to recover single sources and infer skull conductivity values.
In Section~\ref{sec:Implementation}, we present technical details regarding the meshes used in forward simulations and inversion, dipole source spaces, training data, approximation error sampling, and parameter selection procedures in skull conductivity inference. In addition, we describe the used testing data and the different estimates that were computed.
Subsequently, in Subsection~\ref{sec:Results}, we describe our simulated experiments and present the results under different scenarios to demonstrate the proposed approaches. Finally, we discuss further extensions and applications in Subsection~\ref{sec:disc}.}

\subsection{Implementation Details}\label{sec:Implementation}
\subsubsection{Meshes}

For our study, we built the two used 3D meshes (a fine and a coarse one) with the help of the MRI data of the so-called ernie subject and {SimNIBS 4} software\footnote{\url{https://simnibs.github.io/simnibs/build/html/index.html}} \cite{PUONTI2020117044}. Four different tissue compartments (scalp, skull, cerebrospinal fluid, and brain) were considered, and 76 electrodes were placed around the head according to the international 10-10 system (see, Fig.\ref{fig:sourcespace}). The fine mesh consisted of 2,103,623 { tetrahedral} elements joined in 377,150 nodes, and the coarse mesh of 743,575 { tetrahedral} elements joined in 136,868 nodes.

\subsubsection{Leadfield Models}
\label{hm}
The leadfield matrices used in this study were constructed with the help of custom made software that exploited Finite Element Method with linear basis functions, as in \cite{Wolters2004}.

First, we created $K=200$ leadfield matrices with skull
conductivity samples $\sigma^{(k)}$ drawn from a bounded Gaussian
distribution $\pi(\sigma)$ with mean $\sigma_*= 0.0103$\,S/m and
standard deviation 0.0035\,S/m. The skull
conductivity distribution ranged from 0.0041\,S/m to 0.033\,S/m according to the values reported in \cite{hoe03,hom95,ayd14}. The rest of the tissue electric conductivity values were 0.43\,S/m for the scalp, 1.79\,S/m for cerebrospinal fluid, and 0.33\,S/m for the brain (gray matter and white matter) \cite{ram06}. We refer
to these leadfield matrices as {\it sample} models, $A(\sigma)$. Out of these, 150 {\it sample} models were used to { estimate} the skull conductivity related approximation error statistics (in Section \ref{sec:MCMC}) and the remaining 50 {\it sample} models were used for the training of the GP (in Section \ref{sec:infer_skull}). 

For the final testing, we created two \emph{accurate} leadfield matrices $A(\sigma)$ with the help of the fine mesh and skull conductivity values 0.0061 S/m and 0.0139 S/m, that were not included in the sample set. In addition, we created a \textit{standard}
model, $A_0$, with the help of the coarse mesh and skull conductivity $\sigma_0=0.0103$ S/m. 

\subsubsection{Source Spaces}

\begin{figure}[ht]
      \centering
          \includegraphics[width=1\columnwidth]{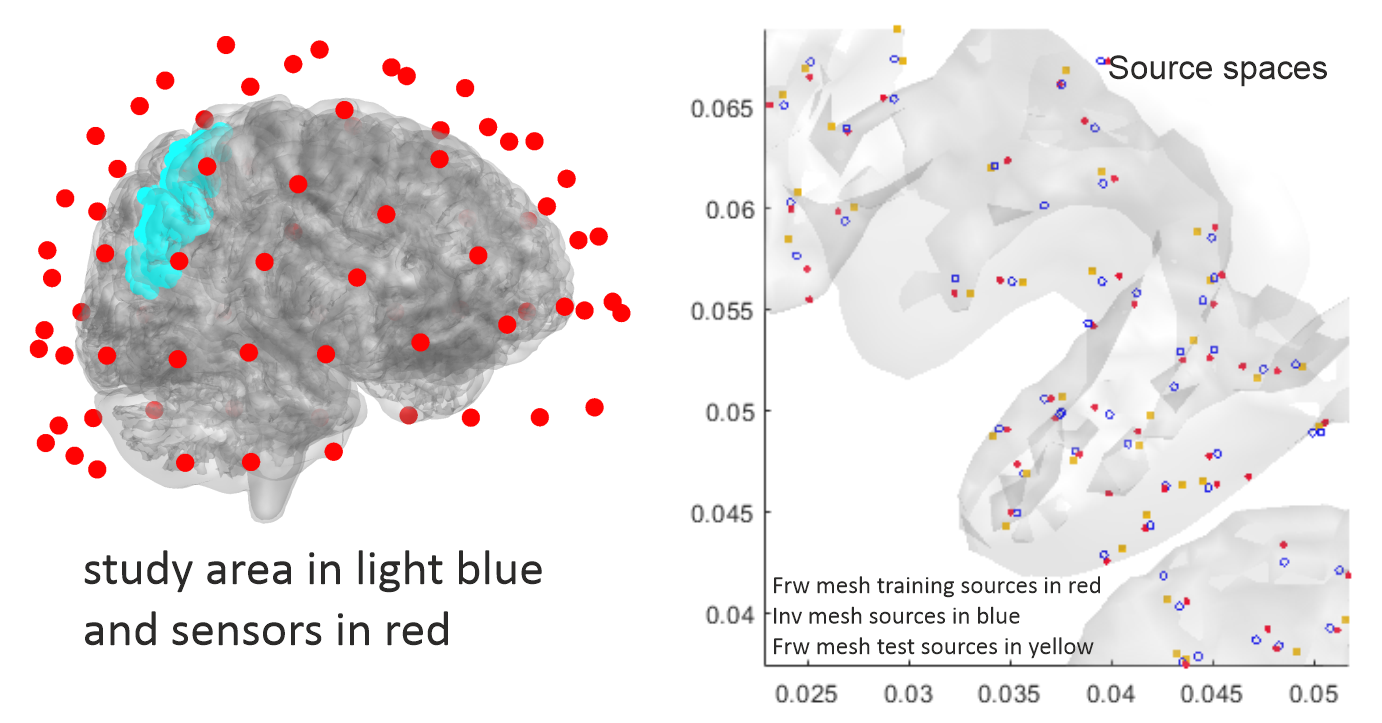}
      \caption{Left: The brain area that was studied and positions of electrodes denoted by red dots. Right: A zoom-in of the area of the brain that was studied to illustrate the candidate source locations. The red dots denote the source locations of the simulated dipoles used for training and production of the approximation error statistics. The blue circles indicate the candidate source locations for the reconstructions. To avoid over-fitting issues, we used different source locations in the final testing marked with yellow circles. The axes are in meters.  }
      \label{fig:sourcespace}
\end{figure}

The dipole source space was restricted to an approximately 30\,mm thick cross sectional area close to the somatosensory area (as shown in the left image of Fig.~\ref{fig:sourcespace}). This area was selected because previous studies had found it suitable for model calibration purposes \cite{Lew09b,Papageorgakis2017,fuc98,ayd14,Antonakakis2020}. {In the right image of Fig.~\ref{fig:sourcespace}, we show a magnification of the region of interest, highlighting the source spaces corresponding to the models considered.}  In particular, the source space of the standard model (in the coarse mesh) is marked with blue circles in the right image of Fig.~\ref{fig:sourcespace}, and it uniformly covers the gray matter of the brain {with 2.5\,mm resolution (distance between neighboring sources)}. The red dots in the right image of Fig.~\ref{fig:sourcespace} indicate the source locations (in the fine mesh) that were used in the estimation of the approximation error statistics, as described in Section \ref{sec:MCMC}. The number of points (source locations) in both source spaces was 560. For the final testing, to avoid over-fitting, we produced observations from simulated dipole sources placed in different locations (in the fine mesh, marked with yellow in the right image of Fig.~\ref{fig:sourcespace}) than the sources that were used to produce the statistics. 

\subsubsection{Approximation Error Statistics and Training Data}
\label{sec:MCMC}

The approximation error statistics were estimated from samples. These samples for each location index $i$ were created by evaluating one of the (randomly chosen) \emph{sample} models $A(\sigma^{(k)})$ and the standard model $A_0$ with a single dipole source $x_i^{(j)}$ as 
\begin{equation}\label{eq:errorSamples}
\varepsilon^{(s)}_i = A(\sigma^{(k)}){x}_i^{(j)} - A_{0}x_i^{(j)}.
\end{equation}
A set of $J=100$ single radial { (with respect to the center of the brain) }dipole samples $x_i^{(j)}$ with amplitudes drawn from a Rayleigh distribution $\pi( |x_i|)|= \text{Rayleigh}(\sqrt{2}\gamma)$ with parameter $\gamma=1.85$ and mode 2.55,  
and 150 different sample leadfield models were used to estimate the approximation error
statistics at each location $i$. The superscripts are as follows:  $s=j+J(k-1)$ where $j=1,\ldots,J$, and $k=1,\ldots, K$, and $s=1,\ldots,S$. 
The mean $\varepsilon_{i*}$ and
covariance matrix $\Gamma_{\varepsilon_i}$ of the corresponding approximation error at location $i$ were
\begin{equation}\label{eq:MeanAndCovariance}
\varepsilon_{i*} = \frac{\sum_{s=1}^{S}{\varepsilon}_i^{(s)}}{S}
\,\,\mbox{and}\,\,\Gamma_{\varepsilon_i}=
\frac{1}{S-1}\sum_{s=1}^{S}(\varepsilon_i^{(s)}-\varepsilon_{i*})(\varepsilon_i^{(s)}-\varepsilon_{i*})^\mathrm{T}.
\end{equation}
Based on these, we evaluated samples for $\alpha_i^{(s)}$ and $\varepsilon_i''^{(s)}$ as $\alpha_i^{(s)} =
W_i^\mathrm{T}(\varepsilon_i^{(s)}-\varepsilon_{i*})$ and
$\varepsilon_i''^{(s)} =
Q_iQ_i^{\mathrm{T}}(\varepsilon_i^{(s)}-\varepsilon_{i*})$, where
$W_i$ and $Q_i$ contained the eigenvectors $1,\dots,p$ and $p+1,\dots,m$ of $\Gamma_{\varepsilon_i}$ respectively (please see Appendix A1 for details). Furthermore, by using these samples we estimated 
numerically $\Gamma_{\alpha_i}$ and the cross-covariance $\Gamma_{\sigma\alpha_i}$, i.e. {
$\Gamma_{\alpha_i} = \frac{1}{S-1} \sum_{s=1}^{S} \alpha_i^{(s)} (\alpha_i^{(s)})^\mathrm{T}$,
 $\Gamma_{\sigma\alpha_i} = \frac{1}{S-1} \sum_{s=1}^{S} (\sigma_i^{(s)} - \sigma_{i*})(\alpha_i^{(s)})^\mathrm{T}$ and  $\sigma_{i*} = \frac{1}{S} \sum_{s=1}^{S} \sigma_i^{(s)}$.}
 
\subsubsection{Selection of $p$ Eigenvectors}

The number of eigenvectors $p$ required to describe the primary error $\varepsilon'$ 
depends on the discrepancies that produce the approximation errors. In general, it is preferable if only few eigenvectors can be used. 
In the current setup, the source amplitude had only a scaling effect on the covariance matrix $\Gamma_{\varepsilon_i}$. 
{Furthermore, we noticed that the numerically estimated covariance is well approximated by the first term of equation (25). Then, since we were looking for a single model parameter value (skull conductivity $\sigma>0$), we set $p=1$ and use only the eigenvector corresponding to the largest eigenvalue. Thus, we had a single error coefficient $\alpha_i\in\mathbb{R}$ to estimate.} 

\subsubsection{Inference of Skull Conductivity from Training Data}
\label{sec:infer_skull}

In this section, we investigate the relationships between the variables $ \{x_i^{(s)},\alpha_i^{(s)},\sigma^{(s)}\}_{s=1}^{N_s}$ for a single dipole location $i$ in order to better understand the skull conductivity estimates that could be obtained from CG (\ref{conditional_conductivity})
and GP (\ref{eq:GP_inference}).

\begin{figure}[ht]
      \centering
          \includegraphics[width=0.99\columnwidth]{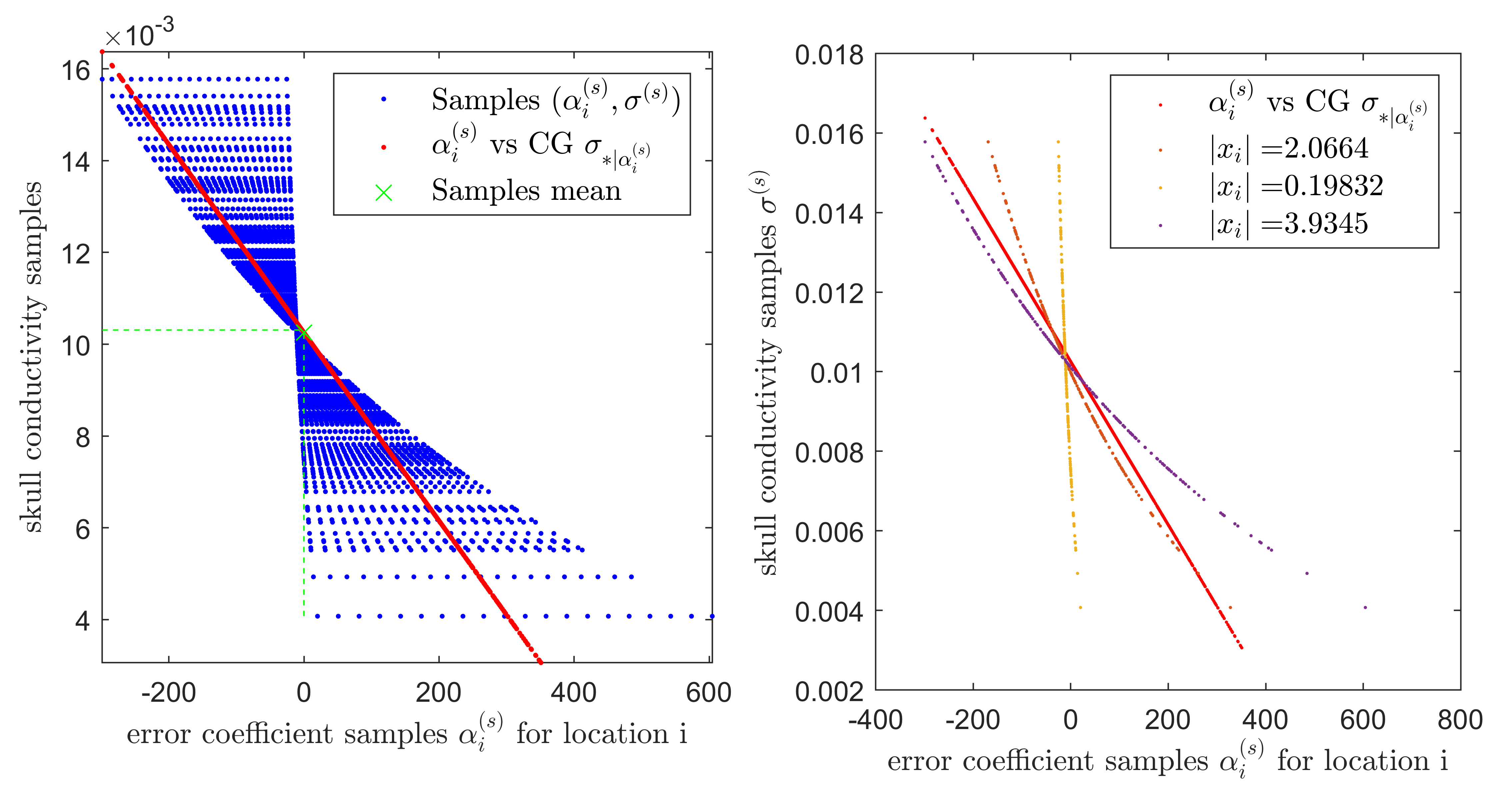}
          \caption{{ In these figures, x-axis shows the sampled $\alpha_i^{(s)}$ and y-axis the corresponding $\sigma^{(s)}$. Left: For a given source location $i$, we visualize the sample pairs $\{\alpha_i^{(s)},\sigma^{(s)})\}$       
          for all the different source amplitudes. The red line shows the CG skull conductivity predictions 
                   $\sigma_{*|\alpha_i} = \sigma_* + \Gamma_{\sigma \alpha_i}\Gamma_{\alpha_i}^{-1}\alpha_i^{(s)}$ (\ref{conditional_conductivity}). The green cross shows the sample mean $\sigma_*$ and the mean of the error coefficient which is zero. Right: Here, we show how the error coefficient $\alpha_i^{(s)}$ is related to the skull conductivity value $\sigma^{(s)}$ given a fixed source amplitude $|x_i|$.} In this case, we can observe monotonic relationships between conductivity $\sigma$ and $\alpha_i$.}
          \label{fig:samples}
 \end{figure}

\paragraph{Conditional Gaussian:}
The red line in Fig.~\ref{fig:samples} shows the conditional skull conductivity predictions for different $\alpha_i$ values using $\sigma_{\mathrm{CG}} = \sigma_* + \Gamma_{\sigma \alpha_i}\Gamma_{\alpha_i}^{-1}\alpha_i^{(s)}$. 
In addition, in the left image of Fig.~\ref{fig:samples} we have visualized the pairs $\{\alpha_i^{(s)},\sigma^{(s)}\}$  for different source amplitudes. The horizontal scatter of the error coefficients $\alpha_i$ for a fixed $\sigma^{(s)}$ shows the effect of varying source amplitudes. Furthermore, the green cross shows the sample mean $\sigma_*$ and the sample mean of the error coefficient. 
We can observe that for skull conductivity values close to the mean $\sigma_*$=0.0103\,S/m (which is equal to the standard conductivity $\sigma_0$), the conditional conductivity predictions from (\ref{conditional_conductivity}) are accurate. Also, we notice that the red curve (CG conductivity prediction curve) passes through the error coefficients that correspond to the mode of the dipole amplitude distribution $\pi(|x_i|)$. Thus, we can expect feasible conductivity predictions only when the source amplitude is close to the mode. When the underlying skull conductivity is further away from $\sigma_*$ the predictions deteriorate. A more detailed explanation for this is given in Appendix A2. 

The right image of Fig. \ref{fig:samples} shows the sample pairs $\{\alpha_i^{(s)},\sigma^{(s)}\}$ for few fixed amplitudes $|x_i|$. We observe monotonic relationships between the conductivity and the error coefficient $\alpha_i$. Also, we can see that when $\sigma>\sigma_0$, the error coefficient $\alpha_i$ is negative, and when $\sigma<\sigma_0$ the error coefficient $\alpha_i$ is positive. Hence, the sign of the error coefficient could indicate whether the true skull conductivity is lower/higher than $\sigma_0$.

\paragraph{Gaussian Process:}
\label{GP_functions}
To allow more complex relationships (instead of just linear) and to take into account the effect of the source amplitude, we employed a Gaussian process to infer the skull conductivity. 
For this purpose, we needed to define the mean
and covariance functions, specified by
hyperparameters. 
In the current problem, the statistics of the underlying (unknown) function $\sigma=f(\alpha_i,x_i)$, $f\sim \mathcal{N}(m(\alpha_i,x_i), K_{\alpha_i,x_i})$, could be approximated based on the expected relationships between the parameters and simulated training data. In particular, the mean function $m(\alpha_i,x_i)$ could be modeled as a polynomial of the following form
\begin{equation}\label{eq:meanGP_ratio}
    m(\alpha_i,x_i)=\sum_{ 0\leq {h} \leq {H}} c_{t}\, \left(\frac{\alpha_i+k}{|x_i|}\right)^{h},
\end{equation}
where $k=w_{i1}^{\mathrm{T}}\varepsilon_{i*}$.  
The elements of the covariance kernel were modeled using the following exponential function
\begin{equation}
    k_{\alpha_i,x_i} = s_f^2 \exp{\left( -\frac{1}{L^2}\begin{bmatrix}
    \frac{\alpha_i}{|x_i|} -\frac{\alpha_i'}{|x_i'|}\end{bmatrix}^{2} 
\right), }
\end{equation}
where the scaling factor $s_f=0.001$ has the same order of magnitude as the standard deviation of the conductivity values (vertical scaling) and 
the length scaling $L$ was set equal to 10 based on the range of $\frac{\alpha_i}{|x_i|}$ values (aiming to avoid under-fitting and over-fitting).    
The selected mean and covariance functions and their parameters encapsulated our beliefs on the data and our understanding of the related physics. A qualitative explanation for these choices can be found in the Appendix A3. 
\begin{figure}[ht]
      \centering
          \includegraphics[width=0.5\columnwidth]{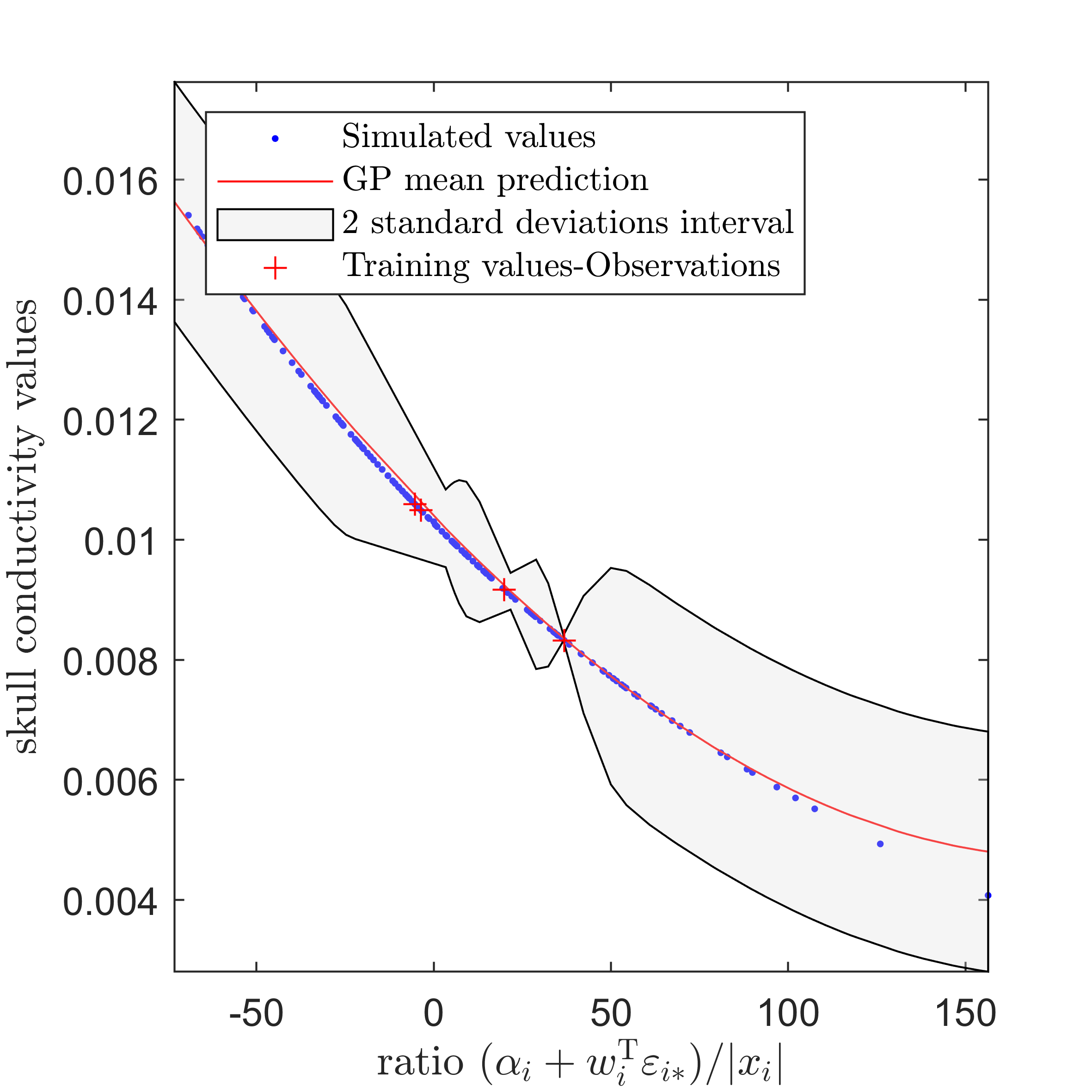}\caption{The blue dots are the target values of the conductivities, the red line is the mean of the GP prediction, the gray color shows the two standard deviation intervals, and the red crosses depict the training data.}\label{fig:GP}
          \end{figure}

In Fig.~\ref{fig:GP}, the blue dots show the simulated (target) values that were not used in the estimation of the $c_t$ coefficient, the red line shows the mean prediction of the Gaussian process regression given by (\ref{eq:meanGP_ratio}), the shaded gray areas represent the (pointwise) two standard deviation limits corresponding to each input, and the red crosses show the observations / training data. Here, we can see that the order $H=2$ is able to describe the relationships between the three parameters. Higher order $H$ did not improve the fit.

In practice to apply the GP, we divided the artificially generated training data set $\{x_i^{(s)},\alpha_i^{(s)},\sigma^{(s)}\}_{s=1}^{N_s}$ into two subsets. The first subset was used to estimate the coefficients $c_t$ through least squares, and the second subset was used later as input-output training data when solving the problem (\ref{eq:GP_inference}).
It is worth noting that the triplets were from additional samples that were not used in the estimation of the approximation error statistics $\Gamma_{\varepsilon_i}$, $\varepsilon_{i*}$ nor $W_i$ in Section~\ref{sec:MCMC}.

\subsubsection{Testing Data and Different Estimates}

For the final testing, the EEG recordings were generated using one of the accurate leadfield models (in the fine mesh) as
\begin{equation}
\label{eq:acc}
    v = A(\sigma_{\mathrm{true}})x + e,
\end{equation}
where the value of $\sigma_{\mathrm{true}}$ was either 0.00601\,S/m or 0.0139\,S/m that were not included in the training sets, and $e$ denotes the additive (white) measurement noise.

\paragraph{Additive Measurement Noise: }
{ Random white noise with zero mean, $e_*=0$, was added to the observations. In particular, the measurement noise was $e=\mathrm{s_c}\;\bar{e}\in\mathbb{R}^{m}$ where $\bar{e}\in\mathbb{R}^{m}$ was sampled from a Gaussian distribution with zero mean and variance one. The scaling parameter $\mathrm{s_c}\in\mathbb{R}^{+}$ was estimated with the help of the chosen signal-to-noise ratio (SNR), defined as $\mathrm{SNR}= 10\log_{10}\frac{\|v\|_2^2}{\|\mathrm{s_c}\;\bar{e}\|_2^2}$ where $\|.\|_2^2$ is the $\ell_2$- norm. 

In the numerical experiments, two levels of SNR were used, 40\,dB and 30\,dB. In the following, the measurement noise covariance was $\Gamma_e=\mathrm{s_c}^2 I_m$.

}

\paragraph{Single Dipole Source Estimates: }

Since we studied only single dipole sources, we used the single dipole scanning algorithm \cite{Fuchs1998,Rodriguez-Rivera2003,Hoeltershinken2024,knosche1997solutions} to estimate the sources. 

\begin{itemize}
    \item When using the {\it standard} leadfield model (in the coarse mesh) with standard skull conductivity $\sigma_0=0.0103$\,S/m, we solved the following optimization problem 
\begin{equation}\label{eq:standardSolution}
   \hat{x}_l\leftarrow \min_{i=1:n} \{ \min_{x_i}\|L_e(v-A_0^i x_i)\|_2^2\},
\end{equation}
where $L_e$ is a matrix square root of $\Gamma_e^{-1}=L_e^T L_e$. In practice, this functional was minimized for each source space node $l$ considering that the dipoles in other locations were zero, and the solution was the dipole source that minimized the residual.

\item When using the proposed Bayesian Approximation Error algorithm, we found the pair that minimized
\begin{equation}\label{eq:ProposedSolution}
(\hat{x}_l,\hat{\alpha}_l)\leftarrow\min_{i=1:n} \{\|L_{\varepsilon^{''}_i+e}(v-A_0^i
\hat{x}_i-\varepsilon_{i*}-w_{i1}{\alpha}_i)\|_2^2+\|L_{\alpha_{i}}{\alpha}_i\|_2^2
\},
\end{equation}
where 
$(\Gamma_{\varepsilon^{''}_i} + \Gamma_e)^{-1} = L_{\varepsilon^{''}_i+e}^{\mathrm{T}} L_{\varepsilon^{''}_i+e}$, and the estimated $\hat{\alpha}_l$ was subsequently used for the skull conductivity estimation. Note that also this dipole source estimate was computed using the {\it standard} leadfield model $A_0$ (i.e. without knowledge of the {\it true} skull conductivity).
\end{itemize}

\subsubsection{Skull Conductivity Estimates}
\label{skullconductivityalgorithms}

Then, the skull conductivity estimation was carried out either by using only the solved $\hat{\alpha}_l$ from the previous step, or the pair $(\hat{\alpha}_l,\hat{x}_l)$. 
\begin{itemize}
    \item CG: The first approach was to use the conditional Gaussian estimation in which the model parameter was solved simply by using the cross-covariance as
\begin{equation}\label{eq:CG_skull}
    \hat{\sigma}_{\mathrm{CG}} = \sigma_* + \Gamma_{\sigma \alpha_l}\Gamma_{\alpha_l}^{-1}\hat{\alpha}_l,
\end{equation}
where $\Gamma_{\alpha_l}$ and $\Gamma_{\sigma,\alpha_l}$ were computed from the samples $\{\alpha_l^{(s)},\sigma^{(s)}\}$.

\item CG+Iter: To improve the above CG solution, we employed the iterative approach presented in Section \ref{sec:iterativealg} (Algorithm \ref{alg:LinearSkull}) that took the estimated source also as an input.
\item GP: Skull conductivity estimation using a Gaussian Process 
\begin{equation}\label{eq:GP_skull}
    \hat{\sigma}_{\mathrm{GP}} =m(\hat{\alpha}_l,\hat{x}_l)+K_{(\hat{\alpha}_l,\hat{x}_l),\mathcal{{D}}}K_\mathcal{D}^{-1}(f_\mathcal{D}-m(\mathcal{D})),
\end{equation}
where $\mathcal{D}=\{\alpha_l^{(s)},x_l^{(s)}\}$, $f_\mathcal{D}=\{\sigma^{(s)}\}$ and the functions as described in Section \ref{GP_functions}.
\end{itemize}

\paragraph{Reference Method: Alternating Single Dipole Scanning: }
{We compared our estimates of dipole source and skull conductivity with the estimates of a reference method based on alternating minimization \cite{Bolte2010,Li2019}. The corresponding estimates were the following}
\begin{equation}\label{eq:altsolution}
  \begin{split}
    \hat{x}_l^{(k+1)}
    & 
   \leftarrow \min_{i=1:n} \{ \min_{x_i}\|L_e(v-A(\sigma^{(k)})^i x_i)\|_2^2\},
    \\
    \hat{\sigma}^{(k+1)}&\leftarrow 
    \min_\sigma\|L_e(v-A(\sigma)^l\hat{x}_l^{(k+1)})\|_2^2,
        \end{split}
\end{equation}
where  $A(\sigma)^l\approx A(\sigma^{(k)})^l+ \frac{\partial A^l}{\partial \sigma}|_{\sigma=\sigma^{(k)}} (\sigma-\sigma^{(k)})$ {where the derivative was evaluated as in \cite{Vauhkonen1997}}. 


\subsection{{Simulated Experiments and Results}}
\label{sec:Results}

We first study the effects of varying source amplitude on source localization accuracy and skull conductivity estimates. Subsequently, we compared the proposed algorithms with a fixed source amplitude. 
In all the test cases, we used single radial dipole sources. We randomly picked 88 source locations (that were not used in the training) from the area of the brain shown in Fig.~\ref{fig:sourcespace}. The EEG measurements $v$ were computed using the {\it accurate} leadfield
model (\ref{eq:acc}) in the fine mesh that had either skull conductivity 0.00601 S/m or 0.0139 S/m (that were not included in the training sets). 

Random white noise was added to the measurements. In Section \ref{ResultsSection1}, a low noise level, $\mathrm{SNR}=40$\,dB, was used in order to study the effects of the varying source amplitude on the results. In Section \ref{ResultsSection2}, a higher noise level, $\mathrm{SNR}=30$\,dB, was used.

All the dipole source reconstructions were carried out using the \emph{standard} leadfield model (in the coarse mesh) that had fixed skull conductivity $\sigma_0=0.0103$\,S/m; the alternating dipole scanning obviously updated the leadfield model during the iterations and used the standard leadfield model only in the initial step. For the model parameter (skull conductivity) estimation, CG+Iter also updated the leadfield model in the iterations.

To evaluate the dipole source reconstructions, we calculated localization errors (Euclidean distances in {millimeters}) between the true and
reconstructed source location. In the following, we denote $X_{\mathrm{st}}$, $X_{\mathrm{BAE}}$, and $X_{\mathrm{alt}}$ as the localization errors of the standard solution (\ref{eq:standardSolution}), BAE solution (\ref{eq:ProposedSolution}), and alternating algorithm (\ref{eq:altsolution}), respectively. Subsequently, we estimated the improvements in source localization $\Delta X$ of the algorithms defined as $X_{\mathrm{st}} - X_{\mathrm{BAE}}$ and $X_{\mathrm{st}} - X_{\mathrm{alt}}$.

\subsubsection{Effects of Varying Source Amplitude in Source Localization Improvement and Skull Conductivity Estimates}
\label{ResultsSection1}

In the first test case, we simulated boundary measurements stemming from 88 radial dipoles (one at a time) with increasing source amplitude (from 0.3 to 4.2, in normalized source amplitude units\footnote{Instead of using the typical single dipole source amplitude values, 10-50 nAm, we used the normalized ones for numerical stability.}), and then computed source reconstructions, using (\ref{eq:standardSolution}) and (\ref{eq:ProposedSolution}), and skull conductivity estimates using the proposed approaches, CG (\ref{eq:CG_skull}), CG+Iter. (Algorithm \ref{alg:LinearSkull}), and GP (\ref{eq:GP_skull}). 

\begin{figure}[ht]
      \centering
          \includegraphics[width=1\columnwidth]{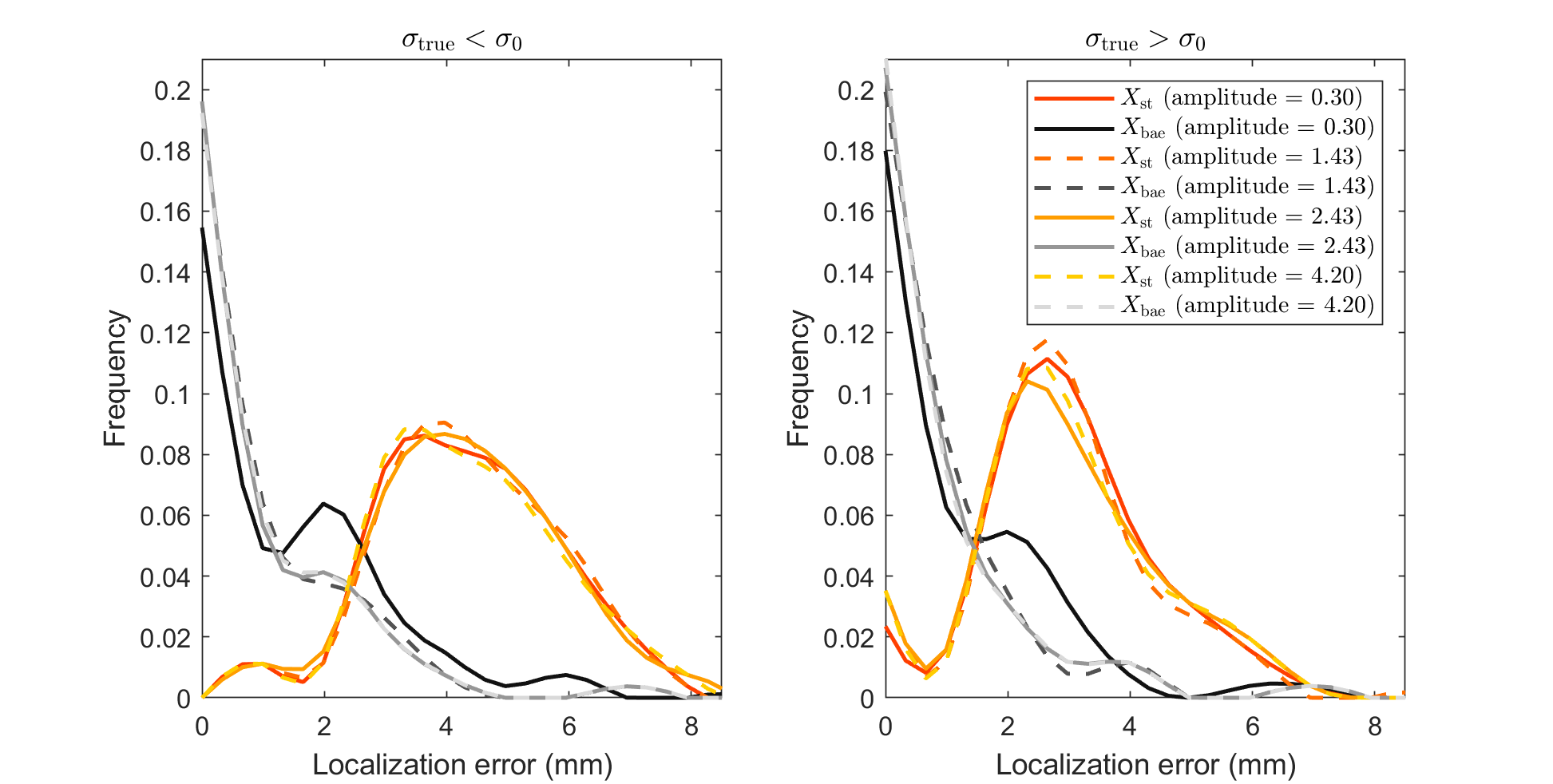}
          \caption{Histogram lines of the localization errors of the standard solution and the BAE solution in the two studied cases, $\sigma_{\mathrm{true}}=0.00601$ S/m and $\sigma_{\mathrm{true}}=0.0139$ S/m, when different source intensities were used. The standard model used in all the reconstructions assumed (erroneously) skull conductivity $\sigma_{0}=0.0103$ S/m.}\label{fig:hists1}
\end{figure}

Figure \ref{fig:hists1} shows {the histograms} of dipole localization errors, $X_{\mathrm{st}}$ and $X_{\mathrm{BAE}}$, of the standard solution (\ref{eq:standardSolution}) and the BAE solution (\ref{eq:ProposedSolution}), respectively, for the four different source intensities. { Each histogram line corresponds to results of 88 simulated sources with a fixed source intensity}. 

As can be seen, the localization errors of the BAE solutions are much more often smaller than the localization errors of the standard solution in all the cases. Moreover, the localization errors do not seem to have any particularly strong trend with respect to the source intensity.  

\begin{figure}[ht]
      \centering
          \includegraphics[width=1\columnwidth]{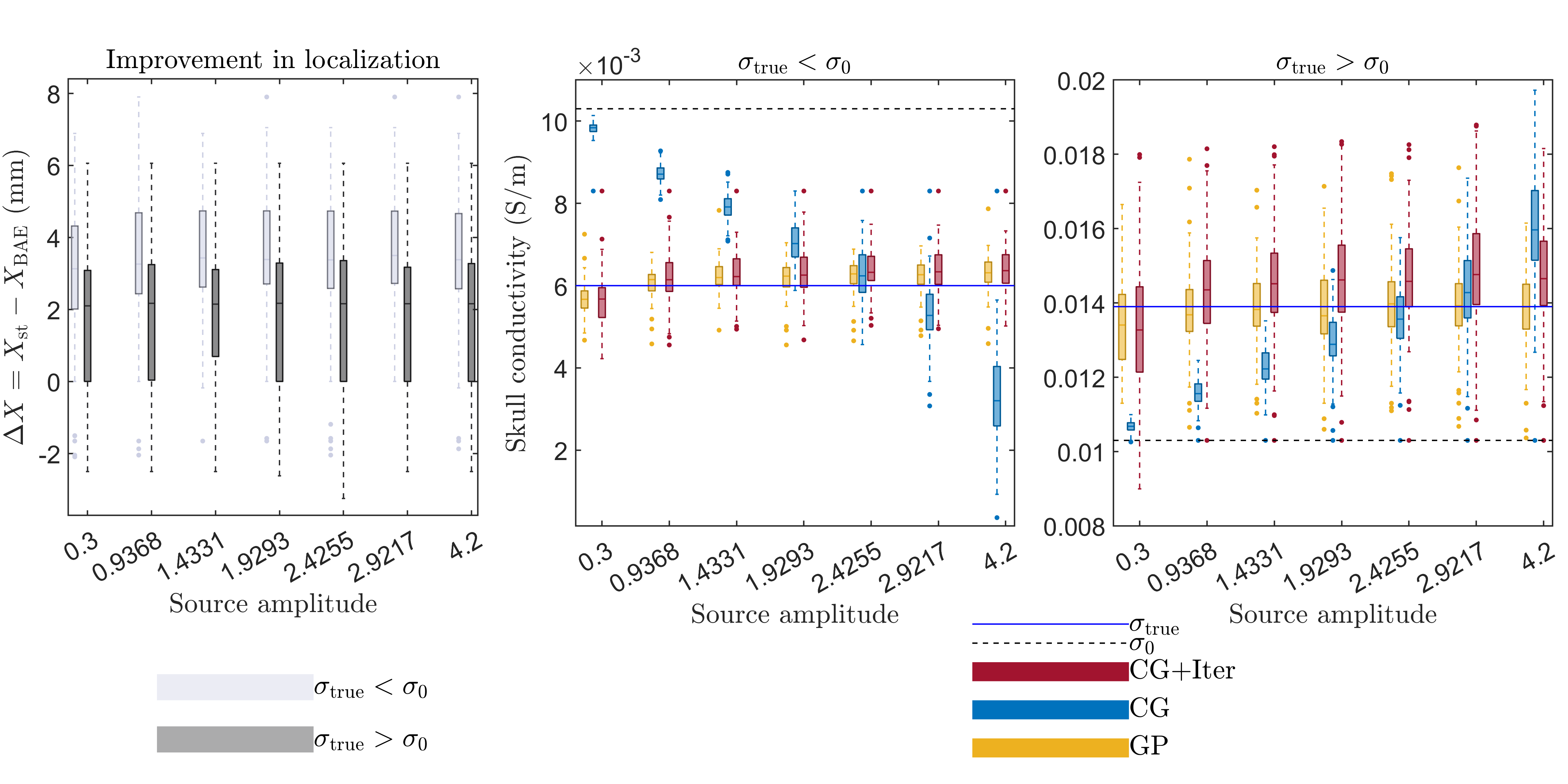}
          \caption{ Comparison of localization improvements $\Delta X$ and skull conductivity estimates across various source amplitudes. The boundary measurements $v$ were simulated by using single dipoles and two testing skull conductivities, $\sigma_{\mathrm{true}}=0.00601$ S/m and $\sigma_{\mathrm{true}}=0.0139$ S/m, while the standard model used in the reconstructions assumed (erroneously) skull conductivity $\sigma_{0}=0.0103$ S/m. The left image illustrates the localization improvements as box plots in these two cases as a function of the varying source amplitude. The images in the middle and right show the estimated skull conductivity values as a function of source amplitude using the three methods (CG, CG+Iter, and GP). The dashed horizontal lines represents the standard conductivity $\sigma_0$ and the blue lines shows the true skull conductivity.}
\label{fig:boxplots}
\end{figure}

In Figure \ref{fig:boxplots}, we show with box plots the source localization improvements ($\Delta X=X_{\mathrm{st}}-X_{\mathrm{BAE}}$, in milli meters) and the estimated skull conductivities in the two test cases as a function of the source amplitude. 
The medians {are} represented by the horizontal lines inside the boxes, the edges of the boxes denote the 25th and 75th percentiles, and the whiskers and outliers indicate the range of values. It is worth emphasizing that the source reconstructions were carried out by first, using the standard leadfield model that assumed (erroneously) skull conductivity value $\sigma_0=0.0103$\,S/m, and second, using the same standard model accompanied with the statistics of the BAE modeling. Higher $\Delta X$ values in the first (left) image of Fig.~\ref{fig:boxplots} indicate greater improvements using the proposed BAE dipole scanning. Light gray refers to the $\sigma_{\mathrm{true}}<\sigma_0$ case and dark gray to the $\sigma_{\mathrm{true}}>\sigma_0$ case. We can see that localization improvements can be achieved using the proposed BAE dipole scanning in both cases, when compared to using the standard dipole scanning. The benefits of BAE are more evident in the case with $\sigma_{\mathrm{true}}=0.00601\,\mathrm{S/m}<\sigma_0$ than in the case with $\sigma_{\mathrm{true}}=0.0139\,\mathrm{S/m}>\sigma_0$.

The second (middle) and third (right) image in Fig.~\ref{fig:boxplots} show box plots of the estimated skull conductivities in the test cases with $\sigma_{\mathrm{true}}=0.00601\,\mathrm{S/m}<\sigma_0$ and $\sigma_{\mathrm{true}}=0.0139\,\mathrm{S/m}>\sigma_0$, respectively. The skull conductivities were estimated using the proposed techniques, CG, CG+Iter and GP, across various source amplitudes. { The horizontal dashed black line represents the standard conductivity $\sigma_0$, and the solid blue line shows the true conductivity used in the test cases}.

Based on these results, the CG-based conductivity estimation performs well only when the source amplitude is close to the mode of the source amplitude distribution $\pi(|x_i|)$ (see Appendix A2 for an explanation). Furthermore, the solution tends to be close to the standard value ($\sigma_0$) when the source amplitude is small. This is because for small source amplitudes, the reconstructed error coefficient ($\alpha_i$) is usually small (due to the scaling effect, see Section \ref{sec:sce}) and thus, favors conductivity values close to $\sigma_0$ (as was expected based on the red curve in Fig.~\ref{fig:samples}). Furthermore, we can see that the skull conductivity estimates of CG for increasing dipole amplitude either decrease (in the $\sigma_{\mathrm{true}}<\sigma_0$ case), or increase (in the $\sigma_{\mathrm{true}}>\sigma_0$ case). Therefore, { we observe here the interaction effect (explained in Sections \ref{sec:overv} and \ref{sec:interact}) based on which a change in the source amplitude can hinder the conductivity estimation with CG, even though, based on the underlying physics, the source amplitude does not depend on the skull conductivity.}

As a first improvement to CG, we proposed an iterative algorithm (Algorithm \ref{alg:LinearSkull}) that takes as an input the estimated source $\hat{x}$ and uses the CG estimated conductivity as an initialization. As can be seen, the accuracy of the iterative algorithm is much better compared to using only CG, and the solution is much less dependent on the source amplitude. 
However, the iterative algorithm is computationally much more effortful, e.g. because of the repetitive computations of the Jacobian matrix. 

As an alternative to CG+Iter., we proposed to use a GP for the conductivity estimation. GP is computationally much less demanding, the effort being essentially the same as in the non-iterative CG. Based on the results, we can observe that GP performs even better than CG+Iter, with medians closer to the true skull conductivity and shorter whiskers.

\subsubsection{Source Localization Improvements and Skull Conductivity Estimates with the Proposed Algorithms}
\label{ResultsSection2}

In the following two case studies, we fixed the source amplitude to 1.3 (normalized units). We estimated the sources and their localization accuracy,  as before. For the model parameter (skull conductivity) estimation, we used only the proposed CG+Iter and GP, as it was shown in the previous section that these two were superior to the simple CG approach. We ran the algorithms for each location with 5 different (measurement) noise realizations at SNR = 30 dB, and we estimated the corresponding localization improvements, $\Delta X=X_{\mathrm{st}}-X_{\mathrm{BAE}}$, and the errors of the conductivity estimates as percentages, $|\Delta \sigma|=100 \times \frac{|\hat{\sigma}-\sigma_{\mathrm{true}}|}{\sigma_{\mathrm{true}}}$. The presented results are the average values of $\Delta X$ and $|\Delta \sigma|$ for each tested location.  
 As a reference method, we computed the source reconstructions and skull conductivity estimates by using the alternating dipole scanning (\ref{eq:altsolution}).

\begin{figure}
    \centering
    \includegraphics[width=1\linewidth]{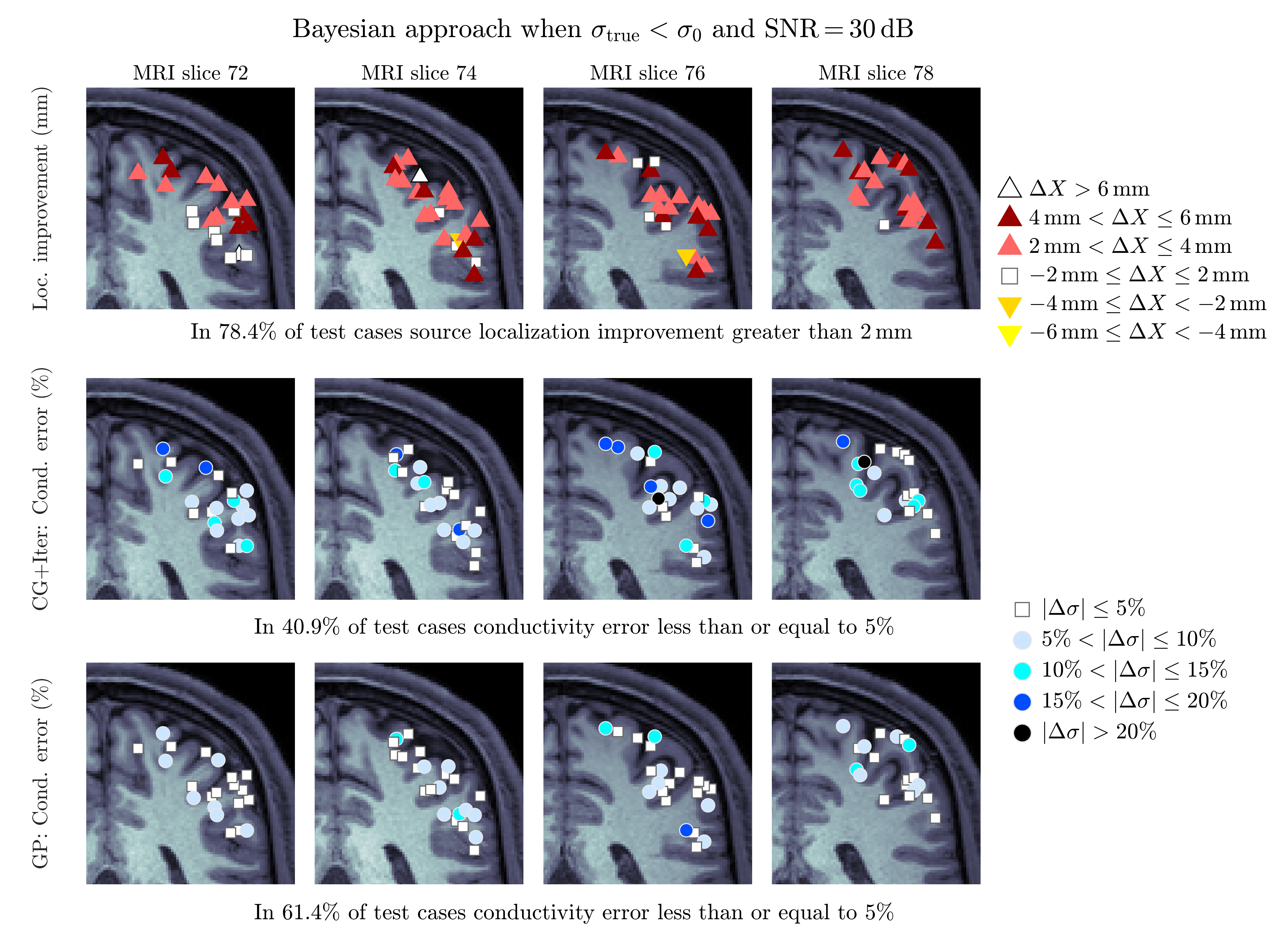}
    \caption{Results of the proposed Bayesian approaches in terms of source localization improvement and conductivity estimation error across {coronal} MRI slices (72, 74, 76, and 78) when $\sigma_{\mathrm{true}}<\sigma_0$ and SNR = 30\,dB. The top row displays localization improvements, with triangles indicating different magnitudes of the improvements (in milli meters). 
    As the conductivity values were estimated together with the sources, both the localization improvements and conductivity estimation errors are presented with the help of the (true) locations of the sources that generated the corresponding EEG testing data. The middle and bottom rows show the errors of the conductivity estimates when CG+Iter. (middle) and GP (bottom) were used. The symbols represent different levels of conductivity errors  (in percentages): white squares denote errors less than 5\%, and circles with shades of blue and black higher skull conductivity estimation errors. Here, GP achieved conductivity errors below 5\% in { 61.4\% of cases, outperforming CG+Iter. that achieved the same only in 40.9\% }of cases.}
    \label{fig:BAE_overestimate}
\end{figure}

Fig.~\ref{fig:BAE_overestimate} depicts the results of the proposed Bayesian approach for the $\sigma_{\mathrm{true}}<\sigma_0$ case applied over selected source locations in MRI slices 72, 74, 76, and 78. 
The top row presents the source localization improvements resulting from the BAE solution with respect to the standard model. 
The white triangles depict the locations in which the source localization improvements were the highest ($\Delta X > 6$\,mm), and the red and orange triangles correspond to smaller localization improvements, ranging from 2\,mm to 6\,mm. In locations marked with white squares, the localization errors were negligibly small ($\leq2$\,mm) discretization considered. The yellow (upside down) triangles depict negative localization improvements (between $-$6\,mm and $-$2\,mm), which means that in these cases the \textit{standard} model (with erroneous skull conductivity value) performed better than the proposed BAE approach. The note below the top row gives the percentage of cases in which the localization improvements were greater than 2\,mm.

\begin{figure}
    \centering
    \includegraphics[width=1\linewidth]{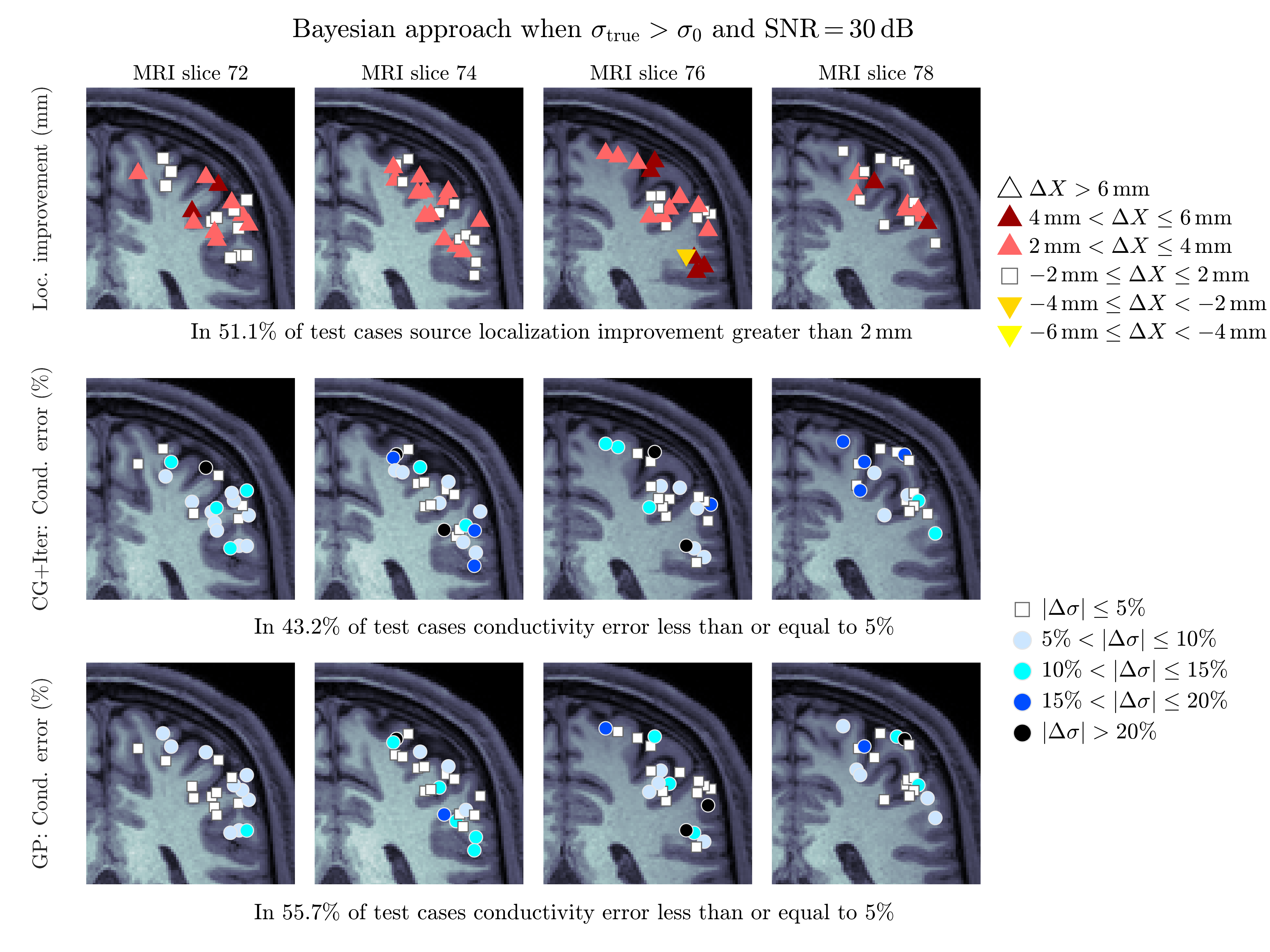}
    \caption{Results of the proposed Bayesian approaches in terms of source localization improvement and conductivity estimation error in the $\sigma_{\mathrm{true}}>\sigma_0$ case. The notations and symbols are as in Fig.~\ref{fig:BAE_overestimate}. }
    \label{fig:underestimatecase_bae}
\end{figure}

\begin{figure}
    \centering
    \includegraphics[width=1\linewidth]{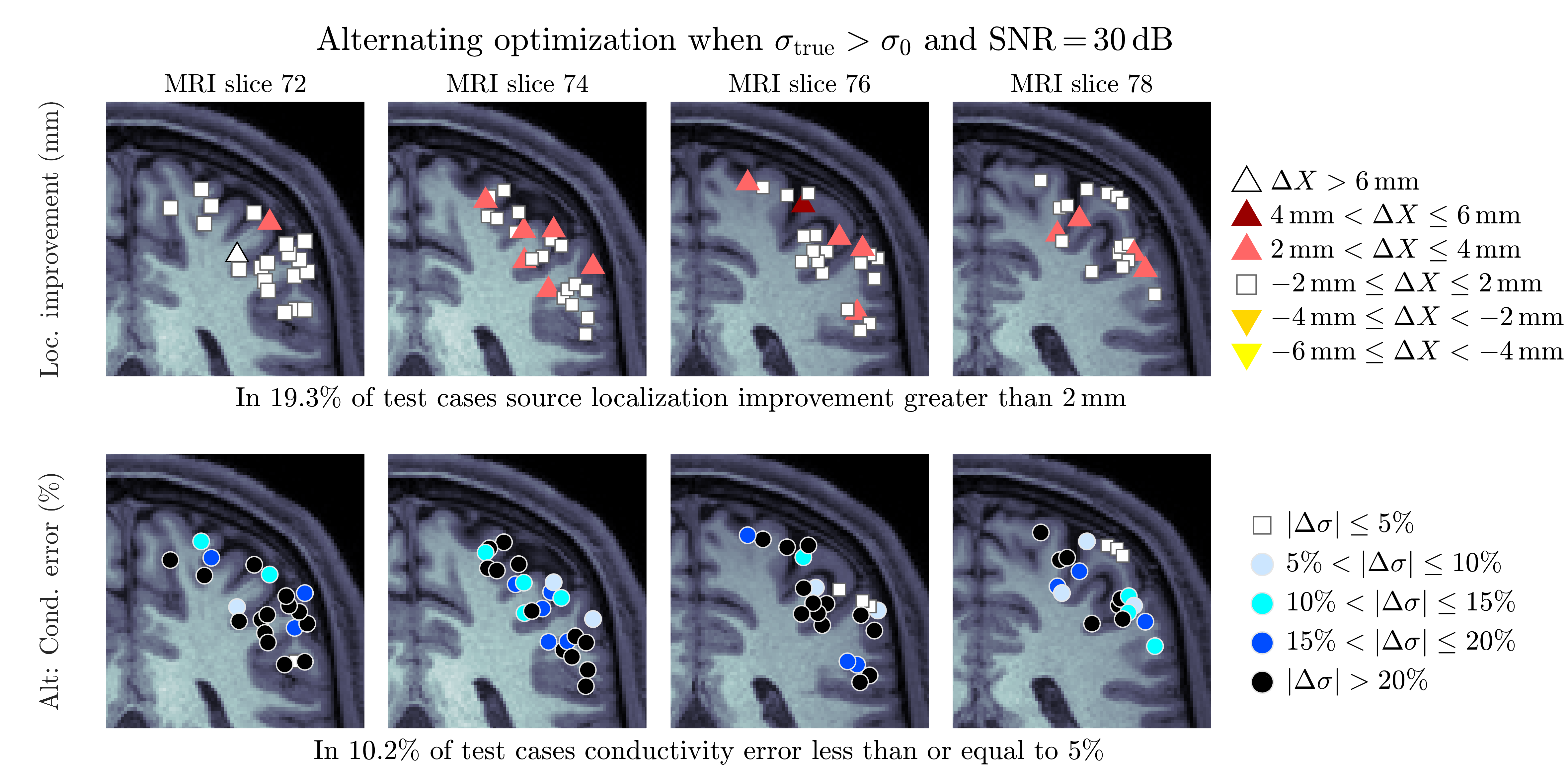}
    \caption{Results of the reference method, the alternating minimization algorithm (\ref{eq:altsolution}), for the $\sigma_{\mathrm{true}}>\sigma_0$ case. The top row shows the localization improvements similarly as the top row of Fig. \ref{fig:BAE_overestimate}, and the bottom row shows the errors of the estimated skull conductivities in percentages similarly as the middle and bottom rows of Fig. \ref{fig:BAE_overestimate}.}
    \label{fig:underestimatecase_alt}
\end{figure}

The middle and bottom rows present data related to conductivity errors, comparing the two different methods, CG+Iter. in the middle row and GP in the bottom row. 
Because the conductivity values were estimated alongside the sources, the conductivity estimation errors are presented with the help of the (true) locations of the sources that generated the corresponding EEG testing data.
The symbols (squares and circles) illustrate different conductivity estimation errors: the white squares indicate the smallest errors ($|\Delta\sigma| \leq 5\%$), and the circles (with different shades of blue and black) indicate higher estimation errors, from 5\% to more than 20\%. 
\begin{figure}
    \centering
    \includegraphics[width=1\linewidth]{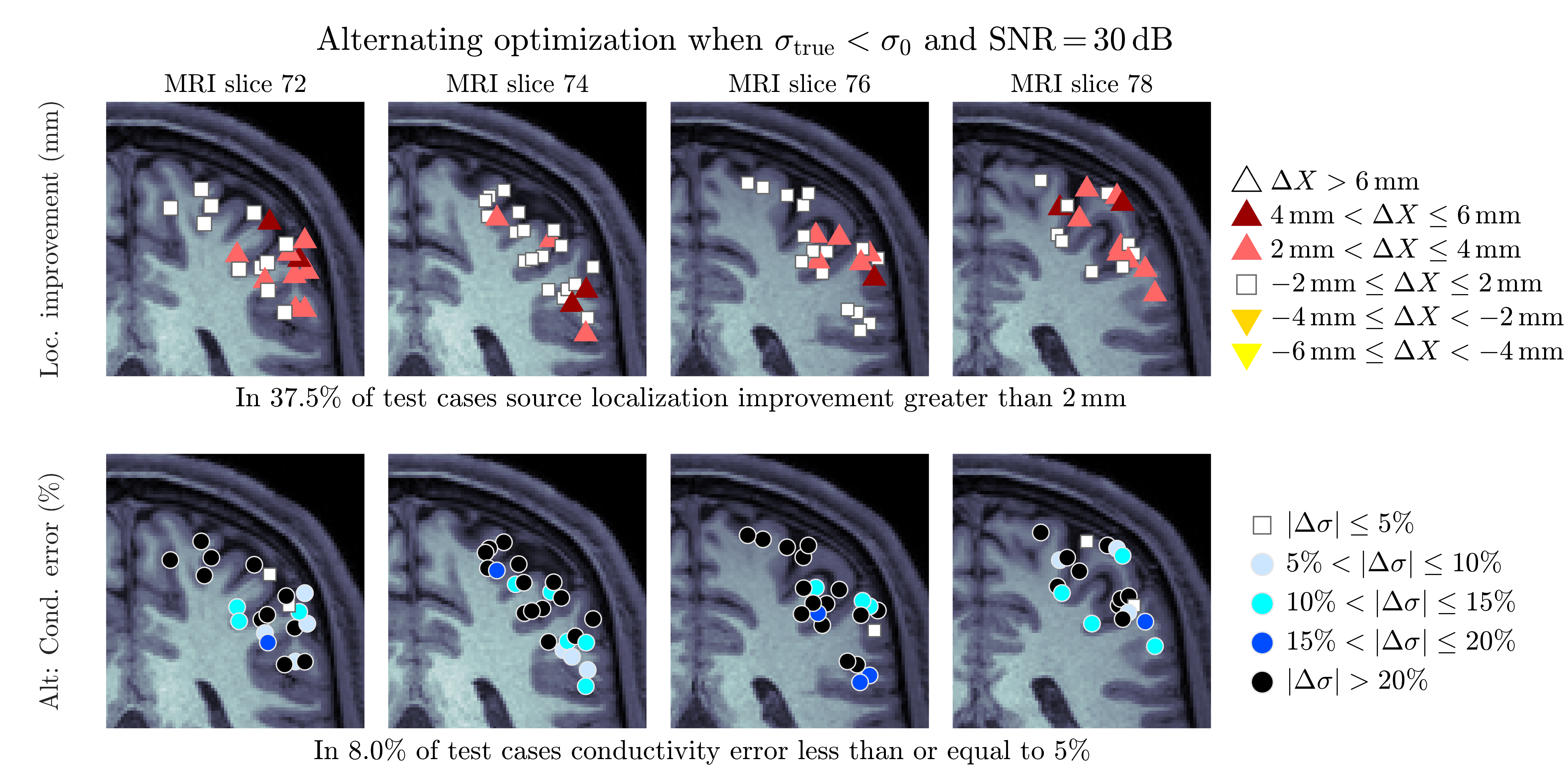}
    \caption{Results of the reference method, the alternating minimization algorithm (\ref{eq:altsolution}), for the $\sigma_{\mathrm{true}}<\sigma_0$ case. The top row shows the localization improvements similarly as the top row of Fig. \ref{fig:BAE_overestimate}, and the bottom row shows the errors of the estimated skull conductivities in percentages similarly as the middle and bottom rows of Fig. \ref{fig:BAE_overestimate}.} 
    \label{fig:alt_overestimate}
\end{figure}
In Fig. \ref{fig:alt_overestimate}, we show the results of the reference method, the alternating dipole scanning set in (\ref{eq:altsolution}), in the $\sigma_{\mathrm{true}}<\sigma_0$ case. The localization improvements and conductivity estimation errors are presented using the same notations and symbols as in Fig. \ref{fig:BAE_overestimate}. Figs. \ref{fig:underestimatecase_bae} and \ref{fig:underestimatecase_alt} present the corresponding Bayesian and reference results in the $\sigma_{\mathrm{true}}>\sigma_0$ case. 

\subsection{Discussion}
\label{sec:disc}

\subsubsection{Source Estimates}

The Bayesian approximation error approach showed substantial improvements in source localization accuracy when compared to both, using the standard model with fixed skull conductivity and the alternating dipole scanning that was used as a reference method. When compared to the standard model, the BAE improved source localization by $\geq2$\,mm in 78.4\% of the tests in the $\sigma_{\mathrm{true}}<\sigma_0$ case and in 51.1\% of the tests in the $\sigma_{\mathrm{true}}>\sigma_0$ case. In the rest of the tests, the performances of the BAE and the standard model were similar; there were only two source locations in which the standard model outperformed the BAE. The alternating dipole scanning performed in most tests equally to the standard model, outperforming the standard model only in 37.5\% of the tests in the $\sigma_{\mathrm{true}}<\sigma_0$ case and in 19.3\% of the tests in the $\sigma_{\mathrm{true}}>\sigma_0$ case.

The improvements provided by the BAE modeling were particularly clear in the case where the true skull conductivity was lower than $\sigma_0$. This is because the EEG measurements, and thus the approximation errors, are non-linearly proportional to the skull conductivity, roughly $\varepsilon \propto 1/\sigma$. 
Therefore, the absolute values of the approximation errors in the $\sigma_{\mathrm{true}}<\sigma_0$ case were (on average) higher than in the $\sigma_{\mathrm{true}}>\sigma_0$ case. 
In the tests where the approximation errors were high (causing poor performance for the standard model), the BAE approach could effectively alleviate the source localization errors with the help of the embedded model-based training statistics of the approximation errors.
The other way around, when the approximation errors were smaller in the $\sigma_{\mathrm{true}}>\sigma_0$ case,  the potential improvements in the localization accuracy that could be gained with the help of the BAE approach were more modest. These findings are in line with our previously published results \cite{Rimpilaeinen2019}. 

\subsubsection{Model Parameter Estimates}
The conditional Gaussian regression-based model parameter estimation did not work robustly for two main reasons. First, the model parameter cannot be solved using only the low-rank estimator for $\varepsilon$, because the solution additionally depends on the source amplitude (as seen in Figs. \ref{fig:samples} and \ref{fig:boxplots}). Second, the model parameter depends non-linearly on $\varepsilon$ (and $\alpha$, as seen in Fig. \ref{fig:samples}) which can cause inaccuracies in the estimates particularly when the \textit{true} conductivity values are relatively far from the linearization point, $\sigma_0$.

Thus, the iterative algorithm that used the CG solution as initialization performed better. The estimates were more accurate because the (estimated) source intensity was taken into account and because the problem was iteratively linearized. The trade-off was the computational burden that significantly increased.

GP offered an effective alternative to the above as it did not require iterations. However, the prerequisite was that suitable mean and kernel functions were needed to describe the relationship between the model parameter and $\varepsilon$. In our case, this relationship could be approximated based on the underlying physics (see Appendix A3). From the tested approaches, GP gave the best model parameter estimates.

The skull conductivity estimates of the alternating optimization (\ref{eq:altsolution}) were poor. Furthermore, we want to stress that the computation of a new lead field matrix and Jacobian matrix, both required by this algorithm, are highly demanding steps, which makes the alternating optimization unfavorable. 
In addition, we observed that the alternating optimization was sensitive to the initialization of the conductivity parameter. In some cases, the method did not converge with the initialization $\sigma_0$, and another initialization point close to $\sigma_0$ was required.

{
\subsubsection{General Remarks, Potential and Limitations}

In this work, by applying the dipole scanning approach that assumes strict single-dipole activity, we successfully removed the dependence of the approximation error on the source location. This was achieved by computing location-specific error estimates; in other words, the error depended solely on the dipole source and the skull conductivity at a fixed location. As a result, the modeling error effects could be described using a single eigenvector, and the computational effort and complexity of the estimation process were reduced considerably in the inversion.

Furthermore, the single-dipole scanning approach employed in our demonstrations has been shown to be equivalent to sLORETA which is a well-established and widely used technique for localizing single smooth source activities \cite{Hoeltershinken2024}. This result  enables direct application of the approach to scenarios involving smoother, spatially extended single sources. In particular, visualizing the sLORETA estimate would be equivalent to visualizing a goodness-of-fit distribution defined by $1-\frac{\|v-A_0^{(i)}\hat{x}^i-w_i\hat{\alpha}_i-\varepsilon_{i*}\|_2}{\|v\|_2}$ for $i=1\ldots,n$, where $(\hat{x}_i,\hat{\alpha_i})$ are the estimates obtained from the single-dipole scan. 

We note that the design of prior distributions (e.g. Gaussian, Laplace or hierarchal priors) in the BAE posterior (\ref{eq:posteriorBAE}) of the EEG source imaging problem is not a straightforward task. This due to the fact that the EEG source imaging problem suffers from an inherent limitation, namely the so-called depth-bias \cite{Badia1998} which means that the reconstructed active sources typically locate close to the brain surface, regardless whether the {\textit{true}} source activity is on the brain surface or deeper in the brain. Unfortunately, this depth-bias problem has not yet been fully resolved; one usually tries to overcome it by using either depth weights and/or hyperparameters in the prior distribution \cite{Koehler1996,PascualMarqui1994,PascualMarqui,Fuchs1999,Wagner2000,PalmeroSoler2007,Buchner1997,Gramfort2014,Koulouri2018,Calvetti2018}. In this work, by considering strictly single-dipole activity, we eliminated the need for these weighting schemes and specially designed priors.

Here, we want to acknowledge that low signal-to-noise ratios can impact both primary unknowns and error estimates, and consequently the accuracy of model parameter inference. However, the specific tolerance to noise depends on properties of the inverse problem, such as the number of measurements and the characteristics of the convolution kernel. In this study, we noticed that with SNR = 30 dB we could still recover meaningful reconstructions. In our previous study \cite{Rimpilaeinen2019}, we observed that the benefits of BAE were negligible when SNR = 20 dB. As detailed in Appendix A4, one way to test this would be to compare the noise level of the measurements with the magnitude of the expected approximation errors \cite{kaipio2013}.

One practical observation we made in the current work was that it is preferable to use an average model parameter value (skull conductivity) in the standard model, since it facilitates model parameter inference under Gaussian assumptions and faster convergence when linearized model parameter estimates are later performed. If the model parameter estimation step is not needed, then the choice of the standard model may not always be very critical \cite{Nicholson_2023}. 

Finally, as the BAE approach without model parameter estimation has been used in various (linear and non-linear) problems \cite{Lipponen2011,Nissinen2009,Arridge2006,Kolehmainen2009,Tarvainen2010a,kou16,Mozumder2013,Alexanderian_2024,Candiani_2021,Nicholson_2023}, we would expect that in many similar cases as these the joint inference of the primary unknown and error approximation (\ref{joint_inference}) would be tractable. Subsequently, based on the available training data, the properties (physics) of the problem at hand, and the complexity of the approximation error term, different inference strategies for the model parameters could be formulated. In addition to the CG, and the proposed CG+Iter. and GP, others such as polynomial chaos expansions and equivalent techniques could be tested for the model parameter inference. 

 }


\section{Conclusion and Future Work}
Inverse problems are challenging due to their ill-posed nature, but also because model uncertainties are often present. The lack of access to ground truth data makes it difficult (or impossible) to apply standard and well-known supervised learning techniques for inferring model parameters. Instead, we must rely on meticulously designed simulations to generate sampling distributions for the model parameters, based on evidence, physics, and carefully made assumptions on the problem. This presents a complex challenge, making the task of designing and creating realistic simulations more critical than ever for effective training and testing, and for avoiding such issues as data snooping and overfitting.

In this paper, we employed the Bayesian Approximation Error approach to solve a blind linear inverse problem with a non-linear (unknown) model parameter in the forward model and tested different approaches to estimate both the primary unknown and the model parameter. To do that, we produced model-based training data that gave us samples of the approximation error. We used these samples to numerically estimate the mean and covariance matrix of the approximation errors, and subsequently a subspace spanned by the top eigenvector of this covariance was used to give a low-rank estimate of the approximation errors in the test cases. 

Based on the solved primary unknown and approximation error estimate, we inferred the (unknown) model parameter using three approaches. The first approach was based on conditional Gaussian regression, the second iteratively updated the linearization point, and the third utilized a Gaussian process that was modeled with the help of physics-informed learning. In addition, alternating optimization was used as a reference method. 

As an application, we studied the EEG source imaging problem in which the (forward) leadfield model contained a non-linear unknown model parameter, the skull conductivity. The utilization of the BAE approach clearly improved the source localization accuracy of the imaging. {{The proposed CG+Iter. and GP provided superior estimates for the skull conductivity when compared to the CG estimates. While the GP and CG+Iter. yielded comparable results across different source amplitudes, the GP method offers a practical advantage due to its computational efficiency and robustness against local minima, making it a preferable choice in most scenarios.}}

In the future, the BAE approach with GPs will be tested with experimental EEG data, and the estimated skull conductivities will be compared to results from other (experimental) skull conductivity calibration techniques.
In EEG imaging, the BAE approach could be tested with more complicated priors, such as $\ell_1$ and group-lasso priors, and in scenarios where several tissue conductivities are recovered simultaneously with the source configuration. { However, in these new formulations, the complexity of the modeling error (e.g., existence of higher-order frequencies and thus the requirement to use more eigenvectors to describe it) should be investigated}. Finally, as the proposed developments are not limited to the EEG source imaging, other inverse problems such as kernel estimation in blind deconvolution problems could be studied.

\section*{Acknowledgements}
This was supported by the Research Council of Finland: Flagship of Advanced Mathematics for Sensing Imaging and Modeling (359185) and PROFI 6–TAU Imaging Research Platform (336357).

\section*{ORCID iDs}

Alexandra Koulouri \url{https://orcid.org/0000-0001-7605-3844}
\\
Ville Rimpil\"ainen \url{https://orcid.org/0000-0001-8679-7898}

{
\section*{Appendix A1: Expressing Approximation Error via Basis Functions}
\label{sec:ErrorViaBasis}

Our aim is to express the approximation error term with the help of a small set of basis functions. We begin with the covariance of the approximation error, defined as
\begin{equation}
\Gamma_{\varepsilon} = \mathbb{E}[(\varepsilon - \varepsilon_*)(\varepsilon - \varepsilon_*)^{\mathrm{T}}],
\label{eq:cov}
\end{equation}
where $\varepsilon_*$ is the mean of the approximation error. Performing eigenvalue decomposition of $\Gamma_{\varepsilon}$ yields
\begin{equation}
\Gamma_{\varepsilon} = \sum_{k=1}^{m} \lambda_k w_k w_k^{\mathrm{T}} = W \Lambda W^{\mathrm{T}},
\label{eq:eigendecomp}
\end{equation}
where $\lambda_k \geq 0$ are the eigenvalues, and $w_k \in \mathbb{R}^m$ are the corresponding orthonormal eigenvectors (principal components) \cite{Nissinen2011a,kaipio2013}.

This decomposition allows us to express the centered approximation error $\varepsilon - \varepsilon_*$ as a linear combination of the eigenvectors:
\begin{equation}
\varepsilon - \varepsilon_* \in \mathrm{span}\{w_1, \ldots, w_m\}.
\label{eq:span}
\end{equation}
In practice, we can write the error as 
\begin{equation}
\varepsilon = \varepsilon_* + \varepsilon' + \varepsilon'',
\label{eq:split}
\end{equation}
where $\varepsilon' = \sum_{k=1}^{p} \alpha_k w_k$ captures the dominant part of the variation, and $\varepsilon'' = \sum_{j=p+1}^{m} \beta_j w_j$ is the residual component. Both components are zero-mean:
\begin{equation}
\mathbb{E}[\varepsilon'] = \mathbb{E}[\varepsilon''] = 0.
\label{eq:zeromean}
\end{equation}

The coefficients $\alpha_k$ and $\beta_j$ are obtained through inner products:
\begin{equation}
\alpha_k = \langle \varepsilon - \varepsilon_*, w_k \rangle, \quad \beta_j = \langle \varepsilon - \varepsilon_*, w_j \rangle.
\label{eq:coeffs}
\end{equation}
Thus, the approximation error can be compactly represented using a reduced basis $W^p = [w_1, \ldots, w_p] \in \mathbb{R}^{m \times p}$ as
\begin{equation}
\varepsilon' = W^p \alpha, \quad \alpha = (\alpha_1, \ldots, \alpha_p)^{\mathrm{T}} \in \mathbb{R}^p.
\label{eq:lowrank}
\end{equation}
The residual term can be written as
\begin{equation}
\varepsilon'' = Q \beta = Q Q^{\mathrm{T}} (\varepsilon - \varepsilon_*),
\label{eq:residual}
\end{equation}
where $Q = [w_{p+1}, \ldots, w_m] \in \mathbb{R}^{m \times (m - p)}$ spans the orthogonal complement. The variances of the principal coefficients are
\begin{equation}
\mathrm{var}(\alpha_k) = \lambda_k, \quad \text{for } k = 1, \ldots, p.
\label{eq:variances}
\end{equation}
}

\section*{Appendix A2: {Predicting Skull Conductivity $\sigma$ under Linear Approximations}}\label{Appendix_1}
In this paper, we studied the conditional Gaussian $\pi(\sigma|\alpha_l)$ and the MAP estimates $\sigma_{\mathrm{MAP}}=\max_{\sigma>0}\log\pi(\sigma|\alpha_l)$. We observed that these estimates were accurate only when the true source amplitude at location $l$ was close to the mode of the distribution of the source amplitude $|x_{l}|$. Here, the source is described by $\mathrm{x}_l=|x_l| \Vec{n}$, considering a fixed orientation $\Vec{n}$.

In this section, we explain why the estimates performed well only under these circumstances. First, we try to derive an explicit expression for the posterior $ \pi(\sigma|\alpha_l)$ using sampling distributions and a linear approximation for the \emph{approximation error}. 

We start with 
$$ \pi(\sigma|\alpha_l) \propto\pi(\sigma)\pi_{|x_l|}(\alpha_l|\sigma)=\pi(\sigma)\int\pi(\alpha_l|\sigma,|x_l|) \pi(|x_l|) \; \mathrm{d}|x_l|.$$
The conditional probability can be approximated as $\pi(\alpha_l|\sigma,x_l)\approx\delta(\alpha_l-G_l(\sigma)|x_l|-c)$, where $\delta(.)$ is the Dirac-delta function and $G_l$ is a function of $\sigma$. 

For small perturbations $\sigma-\sigma_*$, the function $G_l(\sigma)$ can be estimated with the help of the 1st order Taylor series  of the \emph{approximation error} around $\sigma_*=\sigma_0$. In particular, we can write  $$\varepsilon_l=w_{l1}\alpha_l+\varepsilon_{l*}=(\sigma-\sigma_*)\frac{\partial A^{l}(\sigma_*)}{\partial \sigma }\Vec{n}|x_l|, $$ where $\frac{\partial A^{l}(\sigma_*)}{\partial \sigma }\in\mathbb{R}^{m\times 3}$. Now we can multiply both sides with the eigenvector $w_{l1}^\mathrm{T}\in\mathbb{R}^{m}$. This results in
$$\alpha_l+c=(\sigma-\sigma_*)w_{l1}^{\mathrm{T}}\frac{\partial A^{l}(\sigma_*)}{\partial \sigma }\Vec{n}|x_l|,$$
where $c=w_{l1}^\mathrm{T}\varepsilon_{l*}\in\mathbb{R}$. Hence, we can define
$$G_l(\sigma)=(\sigma-\sigma_*)w_l^\mathrm{T} \frac{\partial A^{l}(\sigma_*)}{\partial \sigma}\Vec{n}.$$
Also, we can write
$$|x_l|=G^{-1}_l(\sigma)(\alpha_l+c)=\frac{k^{-1}_l(\alpha_l+c)}{\sigma-\sigma_*},$$ where $k_l=w_{l1}^\mathrm{T} \frac{\partial A^{l}(\sigma_*)}{\partial \sigma}\Vec{n}\in \mathbb{R}$  and where $k^{-1}_l(\alpha_l+c)$ and $\sigma-\sigma_*$ will have the same sign.

When $x_l\in\mathbb{R}^3$ with \( x_l \sim \mathcal{N}(0, \gamma^2 I_3) \),
the distribution of the amplitude \( |x_l| \) is
$
|x_l| \sim \text{Rayleigh}(\sqrt{2}\gamma)$, i.e. $
\pi(|x_l|) = \frac{|x_l|}{2\gamma^2} \exp\left(-\frac{|x_l|^2}{4\gamma^2}\right)$ with $|x_l| > 0
$. 

Therefore, the posterior becomes $$\pi(\sigma|\alpha_l)=\pi(\sigma)\pi_{|x_l|}(G^{-1}_l(\sigma)(\alpha_l+c)).$$
Here, the distribution $\pi(\sigma)$ 
can be approximated as a uniform distribution in the area around the mean $\sigma_*$ and thus 
$\pi(\sigma|\alpha_l)\propto \pi_{|x_l|}(G^{-1}_l(\sigma)(\alpha_l+c))$.

Based on the previous, the log of  $\pi(\sigma|\alpha_l)$
is $$\log \pi(\sigma|\alpha_l) \propto \log \left( G_l^{-1}(\sigma) \right) - \frac{(G_l^{-1}(\sigma)(\alpha_l + c))^2}{4\gamma^2}.$$ To find the MAP estimate, we solve $\frac{d \log \pi(\sigma|\alpha_l)}{d\sigma}=0$ which gives
 $$\hat{\sigma}_{\mathrm{MAP}}=\sigma_* + k_l^{-1} (\alpha_l + c) \frac{1}{\sqrt{2}\gamma},$$where $\sqrt{2}\gamma$ is the mode of the source amplitude. Therefore, the skull conductivity is predicted based on the estimated $\alpha_l$ and the {\textit{mode}} of the source amplitude (instead of the {\textit{estimated}} amplitude).

Therefore, the CG modeling demonstrates good predictive capabilities only when the source amplitude is close to the mode (as was shown with simulations in Section \ref{ResultsSection1}), and more advanced methods that take into account the estimated source amplitude work better in general cases (as was also shown in Section \ref{ResultsSection1}).


\section*{Appendix A3: Choice of the GP Function for Skull Conductivity Prediction in the EEG Problem} 
\label{sec:cond_error_x}

From Ohm's law, we approximate that the electric potential $v$ is proportional to the electric conductivity and a current source as $v \propto \frac{1}{\sigma}|x|$. Furthermore, we approximate that the approximation error (difference of two potentials), $\varepsilon=w_1 \alpha + \varepsilon_*$, is proportional to these variables as 
\begin{equation}
\alpha + w_1^{\mathrm{T}}\varepsilon_* \propto \frac{1}{\sigma}|x|-\frac{1}{\sigma_0}|x|.    
\end{equation}
From this, we write
\begin{equation}
    \sigma \propto(\frac{\alpha+w_1^{\mathrm{T}}\varepsilon_*} {|x|} + \frac{1}{\sigma_0})^{-1}.
\end{equation}
Now, if we denote $y=\frac{\alpha+w_1^{\mathrm{T}}\varepsilon_*}{|x|}$ and define $\tilde{\sigma}(y)=(y+\frac{1}{\sigma_0})^{-1}$, then by applying Taylor expansion around 0, we get $\tilde{\sigma}(y)= \sum_{i=0}^\infty c_i y^i$.

{
\section*{Appendix A4: Pipeline for Model Parameter Estimation Using the BAE Approach}

In this section, we describe an example pipeline to set up the BAE approach with combined estimation of primary unknowns and model parameters. We wish to emphasize that the implementation details are problem-specific. \\ 

\noindent
1. Choose an approximate (discrete) observation model with model parameter $\sigma_0$. The corresponding model for the inversion is then of the form: $v=A(\sigma_0)x+W^p\alpha+\nu \in \mathbb{R}^m$, where $\nu=e+\varepsilon_*+\varepsilon''$ includes mean of the (additive) measurement noise $e_*$, mean of the approximation error term, and the remainder of the approximation error term. Here, the approximation error has been decomposed as $\varepsilon= W^p\alpha+\varepsilon_*+\varepsilon''$. The BAE approach can be used to take into account unknown/uncertain model parameters and numerical (discretization) errors due to model reduction mappings, for example.\\

\noindent    
2. Measurement noise $e$: Usually in calibration and controlled/repeated experiments the level of noise can be low. If possible, the level of measurement noise should be compared to the approximation error term. {For example, one could expect the BAE approach to improve the results when $\| e_*\|^2 + \mathrm{trace} \, \Gamma_e < \| \varepsilon_*\|^2 + \mathrm{trace} \, \Gamma_{\varepsilon}$ \cite{kaipio2013}. In some cases, when
$e_*(k)^2 +\Gamma_e(k,k) < \varepsilon_*(k)^2 + \Gamma_{\varepsilon_*}(k,k)$ holds even for few values of $k$, the BAE approach might still improve the results \cite{kaipio2013}.} Note that here additive measurement noise has been assumed.\\

\noindent    
3. Model-based learning: 

    \begin{itemize}
    \item Estimate the parameters of the \emph{model approximation error}, $\varepsilon=A(\sigma)x-A(\sigma_0)x$, as follows:
    
    \begin{enumerate}
        \item Draw $S+K$ samples: $x^{(s)}\sim \pi(x)$ and $\sigma^{(s)}\sim \pi(\sigma)$
        \item Estimate \emph{sample models}: $\{A(\sigma^{(1)}),\ldots,A(\sigma^{(S+K)})\}$
        \item Evaluate \emph{approximation error} samples: $\varepsilon^{(s)} = A(\sigma^{(s)}){x}^{(s)} - A_{0}x^{(s)}$
        \item Estimate the mean $\varepsilon_{*}$ and covariance matrix $\Gamma_{\varepsilon}$ using the subset $\{x^{(s)},\sigma^{s},\varepsilon^{(s)}\}_{s=1}^S$ as 

            \begin{equation*}
            \varepsilon_{*} = \frac{\sum_{s=1}^{S}{\varepsilon}_i^{(s)}}{S}
            \,\,\mbox{and}\,\,\Gamma_{\varepsilon}=
            \frac{1}{S-1}\sum_{s=1}^{S}(\varepsilon^{(s)}-\varepsilon_{*})(\varepsilon^{(s)}-\varepsilon_{*})^\mathrm{T}.
            \end{equation*}

        \item Estimate basis functions $W$ with the help of, for example, eigenvalue decomposition of $\Gamma_{\varepsilon} = \Sigma_{k=1}^{m} \lambda_k w_k w_k^{\mathrm{T}}=W\Lambda W^\mathrm{T}$, where $\lambda_1>\ldots>\lambda_m$ and $W=[w_1,\ldots,w_m]$.
        
        \item Select the number of basis functions $W^p=[w_1, \ldots, w_p]$ needed for model parameter inference based on the complexity of the model \emph{approximation error}. The complexity can be reflected in the distribution of the eigenvalues. The choice of $p$ is problem specific, and the analysis of the (semi)analytical expression of the error covariance can be helpful, as presented in this paper. For example, one could check whether $\varepsilon \propto x\sigma$ and whether any high-order terms are needed to describe the variations caused by the model parameter. 
        
        \item Evaluate the samples of $\alpha^{(s)} = (W^p)^\mathrm{T}(\varepsilon^{(s)}-\varepsilon_{*})$ with mean $\alpha_*=0$ and covariance $\Gamma_\alpha=\mathrm{diag}(\lambda_1,\ldots,\lambda_p)$. 
        
        \item The remaining error is $\varepsilon''^{(s)} = QQ^{\mathrm{T}}(\varepsilon^{(s)}-\varepsilon_{*})$, where $Q=[w_{p+1},\ldots,w_m]$ 
        with mean $\mathbb{E}[\varepsilon'']=0$ and covariance $\Gamma_{\varepsilon''}$.
      
    \end{enumerate}

\end{itemize} 

\noindent
4. Design the optimization and, if needed, choose appropriate prior for the primary unknown based on the properties of the problem: \begin{equation*}
    (\hat{x},\hat{\alpha}) \leftarrow \min_{x,\alpha} \left\{ \|\frac{1}{2}L_{\nu}(A_0x+W^p\alpha+\nu_*-v)\|_2^2-\ln \pi(x,\alpha) \right\},
\end{equation*}
where $\nu_*=\varepsilon_*+e_*$ and $\Gamma_\nu=\Gamma_e+\Gamma_{\varepsilon''}$. For $\alpha$, one can often use a zero-mean Gaussian smoothness prior. \\

\noindent
5. Model parameter $\sigma$ inference: This step is carried out with the help of the solved $\hat{\alpha}$ from the previous step. In some cases, one may also need the values of $\hat{x}$ for the inference of $\sigma$. 

\begin{itemize}
    \item For the following cases, it is often sufficient to utilize the conditional Gaussian (CG) approximation,  $\hat{\sigma}_{\mathrm{CG}} = \sigma_* + \Gamma_{\sigma \alpha}\Gamma_{\alpha}^{-1} \hat{\alpha}$ where  $\Gamma_{\sigma \alpha}$ is estimated from samples $\sigma^{(s)}$ and $\alpha^{(s)}$. Feasible model parameter estimates can be expected when
    \begin{enumerate}
        \item Operating close to the linearization point, or in other words, when one can assume that the true value of $\sigma$ is close to the chosen approximate $\sigma_0$ that is used in the standard model and the relationship between $\sigma$ and $\alpha$ is linear (or almost linear).
        \item The interaction effect of the primary unknown $x$ on $\sigma$ is only modest. 
    \end{enumerate}
     
    \item In other cases, the model parameter $\sigma$ should be inferred using both $(\hat{\alpha},\hat{x})$. In the following options, the mapping $(\alpha,x)\rightarrow\sigma$ should be smooth and invertible.
    \begin{enumerate}
        \item Option 1: In some cases, such as in this paper, a nonlinear iterative algorithm can be incorporated and initialized with the CG approximation. Iterative algorithms can be time-consuming and computationally effortful. They do not usually require any training data, but then again, they do not benefit from the support of the training data either.
        \item Option 2: A Gaussian Process can be constructed based on the pre-computed samples $\{x^{(s)},\sigma^{(s)},\alpha^{(s)}\}_{s=S+1}^{K+S}$. The mean and covariance functions can be chosen with the help of the underlying physics of the problems and/or the training data. To perform well, the training data should cover a broad range of $\sigma$ values. GPs can be simple and fast to compute, but require an additional training step.
\end{enumerate}

\end{itemize} 

}


\section*{References}
\bibliography{Manuscript_Final_Arxiv.bbl}

\end{document}